\documentclass[
    aps,
    prd,
    twocolumn,
    superscriptaddress,
    nofootinbib,
    longbibliography
]{revtex4-2}

\usepackage{amsmath}
\usepackage{amssymb}
\usepackage{bm}
\usepackage{graphicx}
\usepackage{xcolor}
\usepackage{hyperref}
\usepackage{orcidlink}

\hypersetup{
    colorlinks=true,
    linkcolor=blue,
    citecolor=blue,
    urlcolor=blue,
    pdftitle={Residual Galactic binary foreground in LISA stochastic gravitational-wave background inference: source power concentration and spectral degeneracy},
    pdfauthor={Ruo-Yu Guan and Yan Wang}
}

\begin{document}

\title{
Residual Galactic binary foreground in LISA stochastic gravitational-wave background inference: source power concentration and spectral degeneracy
}
\thanks{Accepted for publication in \href{https://journals.aps.org/prd/}{Physical Review D}. \copyright\ 2026 American Physical Society. Article DOI: \href{https://doi.org/10.1103/4779-jw6l}{10.1103/4779-jw6l}.}

\author{Ruo-Yu Guan\orcidlink{0009-0000-8680-1762}}
\email[Contact author: ]{ruoyuguan.physics@gmail.com}
\affiliation{National Gravitation Laboratory, MOE Key Laboratory of Fundamental Physical Quantities Measurements, Department of Astronomy and School of Physics, Huazhong University of Science and Technology, Wuhan 430074, China}

\author{Yan Wang\orcidlink{0000-0001-8990-5700}}
\email[Contact author: ]{ywang12@hust.edu.cn}
\affiliation{National Gravitation Laboratory, MOE Key Laboratory of Fundamental Physical Quantities Measurements, Department of Astronomy and School of Physics, Huazhong University of Science and Technology, Wuhan 430074, China}

\date{November 5, 2026}

\begin{abstract}
Galactic compact binaries are expected to form a dominant millihertz
foreground for the Laser Interferometer Space Antenna (LISA), and residual
foreground power can bias stochastic gravitational-wave background (SGWB)
inference or increase its uncertainty. Using the LISA Data Challenge 2A
(LDC2A) injections and the Erebor injection-recovery match, we construct a catalog
residual spectrum from \(0.4\) to \(6.0\,{\rm mHz}\). We distinguish two effects. The first is the departure from Gaussianity of a sum of source signals with independent random phases, quantified by its excess kurtosis; the second is the spectral degeneracy between the residual and the SGWB in the mean binned power. Source discreteness
determines the excess kurtosis of this sum, but instrumental noise and
fiducial SGWB power strongly reduce its magnitude in most bins. We quantify this
degeneracy with a local Fisher information matrix for the SGWB
amplitude \(\Omega_0\) and the residual-power factor \(\beta\). For a
frequency-independent SGWB with \(\Omega_0=10^{-11}\) and residual power
distributed uniformly over the Fourier frequencies within each bin,
marginalizing over \(\beta\) increases the \(\Omega_0\) uncertainty by
\(13.6\%\) with the Gaussian Fourier-amplitude covariance and by \(13.8\%\)
with the finite-source diagonal random-phase covariance. Under the Gaussian
Fourier-amplitude covariance, the largest allowed prior standard deviation
on \(\beta\) that limits this increase to \(10\%\) is \(0.0073\). With this
covariance recomputed at each fiducial amplitude, the uncertainty increase
ranges from \(12.7\%\) to \(20.7\%\) over
\(10^{-13}\leq\Omega_0\leq10^{-10}\). The residual--SGWB spectral degeneracy
persists under alternative distributions of residual power over Fourier
frequencies and diagonal covariances, while omitting the
residual induces a systematic bias in the fitted SGWB amplitude.
\end{abstract}

\maketitle

\section{Introduction}
\label{sec:intro}

The Laser Interferometer Space Antenna (LISA) will observe the combined strain
of many millihertz gravitational-wave signals
~\cite{AmaroSeoane2017LISA}. Galactic compact binaries, especially double white
dwarfs, are expected to be among the most numerous sources. Tens of
millions of binaries may populate the LISA band, but only a fraction will be
individually resolved; the remainder will form an unresolved Galactic
confusion foreground
\cite{Hils1990Galaxy,BenderHils1997Confusion,Nelemans2001GalacticDisk,
Timpano2006Foreground,Ruiter2010Foreground,Korol2022DWD}. The LISA Data
Challenges and the earlier Mock LISA Data Challenges provide simulated data
sets for developing analysis methods, including the LISA Data Challenge 2A
(LDC2A) data set used here to study the source confusion regime~\cite{Babak2008MLDC,Baghi2022LDC}.

Earlier work developed pipelines for resolving and subtracting Galactic
binaries in simulated LISA data
\cite{CornishLarson2003LISASubtraction,2021PhRvD.104b4023Z,
2022PhRvD.106j2004Z}.
Bayesian methods have long been used to resolve overlapping Galactic
binaries
\cite{CrowderCornish2007Foreground,Littenberg2011DetectionPipeline}.
More recent global-fit frameworks jointly model multiple source populations
and instrumental noise
\cite{LittenbergCornish2023GLASS,Strub2024GlobalAnalysis,
Deng2025ModularGlobalFit}. The Erebor analysis of the LDC2A data set produced a large Galactic binary catalog
~\cite{Katz2025Erebor}. Related work has addressed the construction of source
catalogs from global fit posterior samples, for which label switching and a
varying number of sources complicate the association of samples with
individual binaries
~\cite{Johnson2025PETRA}.

The unresolved or imperfectly subtracted Galactic binary foreground is also
relevant to searches for a stochastic gravitational-wave background (SGWB)
~\cite{CapriniFigueroa2018CosmologicalBackgrounds,
Renzini2022SGWBReview}. Simulated Galactic white dwarf foregrounds show
departures from Gaussianity and stationarity across parts of the millihertz
band, motivating data models that account for both properties
~\cite{Buscicchio2025ForegroundGaussianity}. Separating an SGWB from
instrumental noise and astrophysical foregrounds also requires spectral and
detector response information
\cite{AdamsCornish2010SGWBNoise,AdamsCornish2014SGWBForeground,
Boileau2021SpectralSeparation}. Rosati and
Littenberg studied SGWB recovery from LISA global fit residuals and found that
unmitigated residual power and non-Gaussianity can alter upper limits and
produce false detections for some SGWB spectra
~\cite{Rosati2024SGWBResidual}.

In an analysis of LISA observations, the residual foreground is not known in
advance. It can be represented by posterior residual realizations or by an
uncertain foreground model propagated from a global fit into the SGWB analysis
\cite{LittenbergCornish2023GLASS,Johnson2025PETRA,
Rosati2024SGWBResidual}. Injection--recovery associations instead provide the
catalog-based residual spectrum used here to quantify how uncertainty in its
power affects SGWB inference.

We quantify how uncertainty in a Galactic binary residual spectrum constructed
from catalog data propagates into the inferred SGWB amplitude. We use the
LDC2A injection catalogs and the Erebor injection-recovery match to
construct a residual spectrum from injected sources that do not satisfy the
adopted recovery criteria. Let \(\Omega_{\rm GW}(f)\) denote the dimensionless
SGWB energy density per logarithmic frequency interval, \(\Omega_0\) its
amplitude at a reference frequency, and \(\beta\) a dimensionless scaling
factor for the catalog residual power spectrum. A local Fisher information
matrix for binned power measurements then quantifies the joint constraints on
\(\Omega_0\) and \(\beta\)
~\cite{CutlerFlanagan1994ParameterEstimation,Vallisneri2008FisherMatrix}.
In this calculation, the injection-based spectrum supplies the spectral shape
\(R_k\), while \(\beta\) represents uncertainty in its overall power.

We study two effects of the catalog residual. First, to characterize
the departure from Gaussian statistics, we consider a sum of the
residual source signals with independent random phases. Within this
construction, the size of this departure is set by the source
discreteness. It is strongest when a few sources dominate the bin
power, and the instrumental noise and the fiducial SGWB power dilute
it. Second, the residual spectrum overlaps the SGWB spectrum in the
mean binned power, so the two amplitudes are partially degenerate when
they are inferred jointly. We quantify the first effect by the excess
kurtosis of this sum, and the second by a local
Fisher analysis of the mean binned power using three covariances.
Marginalizing over \(\beta\) increases the uncertainty
on \(\Omega_0\), whereas omitting the residual biases the inferred
amplitude.

Using an orbit-averaged long-wavelength Michelson \(X\) response over the
\(0.4\)--\(6.0\,{\rm mHz}\) band, we find that the excess kurtosis of this
sum is strongly diluted in most bins. For a frequency-independent SGWB
with \(\Omega_0=10^{-11}\), freely marginalizing over \(\beta\) increases the
full-band \(\Omega_0\) uncertainty by \(13.6\%\) with the Gaussian
Fourier-amplitude covariance and by \(13.8\%\) with the finite-source
diagonal random-phase covariance. Alternative residual frequency
distributions and covariances change the size of the uncertainty increase
but do not remove the residual--SGWB spectral degeneracy.

This paper is organized as follows.
Section~\ref{sec:data} describes the catalog data, residual construction,
source power convention, and source discreteness.
Section~\ref{sec:covariance} derives the covariance of binned Fourier power
and discusses how instrumental noise and SGWB power dilute the excess kurtosis.
Section~\ref{sec:model}
introduces the binned SGWB model and Fisher analysis.
Section~\ref{sec:finite_bin_results} presents the all-bin and
frequency-subset results, and Sec.~\ref{sec:discussion} discusses the main
conclusions and limitations.
Appendix~\ref{app:additional_checks} collects additional covariance,
binning, and frequency-selection results; Appendices~
\ref{app:random_phase_moments} and \ref{app:fisher_prior} give the
supporting analytic derivations; and Appendix~
\ref{app:numerical_conventions} documents the catalog matching, detector
response, and numerical conventions.

\section{Catalog data, residual construction, and source discreteness}
\label{sec:data}

We use the detached and interacting Galactic binary injection catalogs from
the LDC2A data set
~\cite{Baghi2022LDC,Baghi2022LDCZonedo}, together with the Erebor
injection-recovery match from the Galactic binary global fit
~\cite{Katz2025Erebor,Katz2024EreborLDC2A}. The injection catalogs define the
reference population, and the match supplies the reported confidence and
overlap used to classify the injected sources.

Although the LDC2A release contains the information required to generate
time-domain data, we do not use the time-series values themselves. We use
catalog frequencies and frequency derivatives, strain amplitudes,
inclinations, sky positions, and polarization angles, together with the LDC
orbit model to compute the orbit-averaged detector response. The sample count and cadence
of the LDC2A observation determine the Fourier frequency grid. Instrumental
noise and the SGWB contribution are modeled by analytic spectral densities
evaluated on this grid.

\subsection{Catalog data and frequency binning}

We divide the range
\(0.4\,{\rm mHz}\leq f<6.0\,{\rm mHz}\) into 560 equal-width bins with
\(\Delta f=10^{-5}\,{\rm Hz}\). This band spans the transition from the
low-frequency foreground, where many Galactic binaries contribute to each bin,
to a sparser regime in which recovered sources account for most of the
Galactic binary power. The lower and upper boundaries exclude the densest
low-frequency foreground and the sparse high-frequency tail, respectively.
The adopted bin width preserves the broad frequency dependence of the catalog
residual while containing many Fourier frequencies of the one-year LDC2A
observation.

Injected sources are assigned to analysis bins according to their intrinsic
gravitational-wave frequencies. Their one-year intrinsic frequency drifts and
annual Doppler ranges are narrower than the adopted bin width, although
sources near a bin boundary can transfer power to an adjacent bin. The
corresponding frequency intervals and the resulting inter-bin power transfer
are quantified in Appendix~\ref{app:numerical_conventions}.

An injected Galactic binary is classified as confidently recovered when
\begin{equation}
    C_{\rm rec}\geq0.9\, ,
    \qquad
    {\cal O}_{\rm best}\geq0.9\, ,
\end{equation}
where \(C_{\rm rec}\) is the recovery confidence reported as \(C\) in the
Erebor injection--recovery match and \({\cal O}_{\rm best}\) is the reported
best injection--recovery overlap. We write the confidence as \(C_{\rm rec}\)
to distinguish it from the source discreteness \(C_2\) defined below.
Each accepted match row is associated with a unique LDC2A source using the
injected source parameters provided in the catalogs. Sources satisfying both
recovery criteria are classified as recovered and excluded from the catalog
residual; the remaining injected sources define the residual population.

This construction represents the foreground power remaining after confidently
recovered sources are removed at the catalog level. It does not propagate
posterior uncertainty in recovered-source parameters or residual power from
imperfect waveform subtraction.

Instrumental noise is modeled using the optical metrology and acceleration
noise levels adopted for the LISA sensitivity
curve~\cite{Robson2019Sensitivity}, converted to equivalent incident strain for the long-wavelength Michelson
\(X\) channel. We denote its binned power by
\(P_{{\rm inst},k}\); the corresponding spectral density and Fourier-frequency
sum are specified in Appendix~\ref{app:numerical_conventions}.

\subsection{Source strain power weights and residual spectrum}

We assign each injected source a nonnegative equivalent incident strain power
\(p_i\). It is defined from a long-wavelength Michelson \(X\) response,
averaged over the orbital phase of the equal-arm LDC orbit and divided by the
sky- and polarization-averaged response of one Michelson channel. The detector
geometry is given in Appendix~\ref{app:numerical_conventions}.

For a source with catalog strain amplitude \({\cal A}_i\) and inclination
\(\iota_i\), the plus and cross strain amplitudes follow the LDC
Galactic-binary convention~\cite{Babak2008MLDC,Baghi2022LDCZonedo}:
\begin{equation}
    h_{+,i}
    =
    {\cal A}_i
    \left(1+\cos^2\iota_i\right)\, ,
    \qquad
    h_{\times,i}
    =
    2{\cal A}_i\cos\iota_i\, .
\end{equation}
At orbital phase \(\varphi\), the Michelson \(X\) power contribution after
averaging over source phase is
\begin{equation}
    p_i^X(\varphi)
    =
    \frac{1}{2}
    \left[
        \left(F^X_{+,i}(\varphi)h_{+,i}\right)^2
        +
        \left(F^X_{\times,i}(\varphi)h_{\times,i}\right)^2
    \right]\, ,
\end{equation}
where \(F^X_{+,i}\) and \(F^X_{\times,i}\) are the antenna factors for the
catalog sky position and polarization angle
~\cite{Cutler1998AngularResolution,CornishRubbo2003LISAResponse}. For one
Michelson channel, the long-wavelength response for either polarization,
averaged over sky position and polarization angle, is
\begin{equation}
    {\cal R}_{\rm LW}^{X}
    =
    \left\langle\left(F_+^X\right)^2\right\rangle
    =
    \left\langle\left(F_\times^X\right)^2\right\rangle
    =
    \frac{3}{20}\, .
    \label{eq:lwa_x_response}
\end{equation}
We therefore define the catalog power weight as
\begin{equation}
    p_i
    =
    \frac{\left\langle p_i^X(\varphi)\right\rangle_{\rm orb}}
         {{\cal R}_{\rm LW}^{X}}\, ,
    \label{eq:response_normalized_source_power}
\end{equation}
where \(\langle\cdot\rangle_{\rm orb}\) denotes the average over one LISA
orbit.

The division by \({\cal R}_{\rm LW}^{X}\) places the catalog source powers, the
single-channel instrumental noise, and the SGWB spectrum in the same equivalent
incident strain convention. The orbital average retains the dependence of the
Galactic catalog on sky position without constructing a time-dependent
likelihood. The response weighting itself does not include Doppler frequency
modulation or finite-arm transfer functions. Under this convention, \(p_i\) contributes to a binned strain
power sum and has units of strain squared.

For each frequency bin \(k\), we define the residual power and the sum of
squared residual source powers as
\begin{equation}
    R_k
    =
    \sum_{i\in k,\ {\rm residual}} p_i\, ,
    \label{eq:residual_template}
\end{equation}
\begin{equation}
    S_{2,k}^{\rm res}
    =
    \sum_{i\in k,\ {\rm residual}} p_i^2\, .
\end{equation}
The superscript \({\rm res}\) in \(S_{2,k}^{\rm res}\) denotes the residual
source population. The values \(R_k\) define the catalog residual spectrum
used in the SGWB analysis. The total injected Galactic binary power in bin
\(k\) is
\begin{equation}
    P_{{\rm inj},k}
    =
    \sum_{i\in k,\ {\rm injected}}p_i\, .
\end{equation}

For comparison with the catalog residual, we denote the binned SGWB
contribution by \(P_{{\rm SGWB},k}=\Omega_0T_k(\alpha)\), where
\(\alpha\) is the SGWB spectral index and \(T_k(\alpha)\) is the binned
strain-power template per unit \(\Omega_0\).
For \(\alpha=0\), the underlying strain spectral-density template is
proportional to \(f^{-3}\). The SGWB model and the construction of
\(T_k(\alpha)\) are specified in Sec.~\ref{sec:model}. Thus \(R_k\),
\(P_{{\rm inst},k}\), and \(\Omega_0T_k(\alpha)\) are all expressed as
equivalent incident strain power and have the same units as \(p_i\).

The SGWB inference model scales the catalog residual spectrum by a
dimensionless factor \(\beta\), with \(\beta_{\rm fid}=1\) corresponding to
the catalog residual constructed above. Uncertainty in this normalization is
represented by a prior on \(\beta\), as described in
Sec.~\ref{sec:model}.

Table~\ref{tab:analysis_settings} summarizes the catalog, frequency, response,
and model settings.

\begin{table*}[t]
\centering
\caption{Summary of the settings used in the catalog analysis.}
\label{tab:analysis_settings}
\small

\begin{ruledtabular}
\begin{tabular}{p{0.20\textwidth}p{0.34\textwidth}p{0.38\textwidth}}
Quantity & Value & Comment \\
\hline
Catalog data &
LDC2A injections and the Erebor injection-recovery match &
Detached and interacting Galactic binary catalogs; injected sources that
do not meet the recovery criteria define the catalog residual \\

Frequency binning &
\(0.4\)--\(6.0\,\mathrm{mHz}\), \(\Delta f=10^{-5}\,\mathrm{Hz}\) &
560 equal-width bins; 315 or 316 positive Fourier frequencies enter each
binned-power sum \\

Recovery criteria &
\(C_{\rm rec}\geq0.9\), \(\mathcal{O}_{\mathrm{best}}\geq0.9\) &
Confident injection--recovery association in the Erebor match \\

Power and noise convention &
orbit-averaged long-wavelength Michelson \(X\) source powers;
single-channel instrumental noise using published LISA noise
levels~\cite{Robson2019Sensitivity} &
Equivalent incident strain power using
\({\cal R}_{\rm LW}^{X}=3/20\) \\

Observation and Fourier frequencies &
\(T_{\rm obs}=31\,536\,000\,{\rm s}\), cadence \(5\,{\rm s}\) &
Positive frequencies \(f_j=j/T_{\rm obs}\); the time-series values are
not used \\

Bin counts &
560 bins in total; 487 with \(R_k>0\) &
All 560 bins enter the full Fisher information matrix; residual statistics use the
487 bins with nonzero residual power \\

Frequency subsets &
\(N_{\rm bin}=6,12,24,36\) &
\(N_{\rm bin}\) is the number of bins in each cumulative subset of the
487 bins with nonzero residual power, ranked by the magnitude of the diluted
excess kurtosis, \(|\gamma_{2,\rm tot,k}|\), defined in Sec.~\ref{sec:covariance} \\

Model settings &
fiducial \(\Omega_0=10^{-11}\), \(\alpha\in\{-2,0,2\}\),
\(\beta_{\rm fid}=1\);
covariance diagonal in frequency bins and held fixed within each local
Fisher calculation &
Gaussian Fourier-amplitude baseline, with finite-source random-phase and
fixed-residual diagonal comparisons; Gaussian prior standard deviation
\(\sigma_\beta\) for \(\beta\);
for \(\alpha=0\), the all-bin calculation also spans
\(10^{-13}\leq\Omega_0\leq10^{-10}\) \\
\end{tabular}
\end{ruledtabular}
\end{table*}

\begin{figure*}[t]
    \centering
    \includegraphics[
        width=0.95\textwidth
    ]{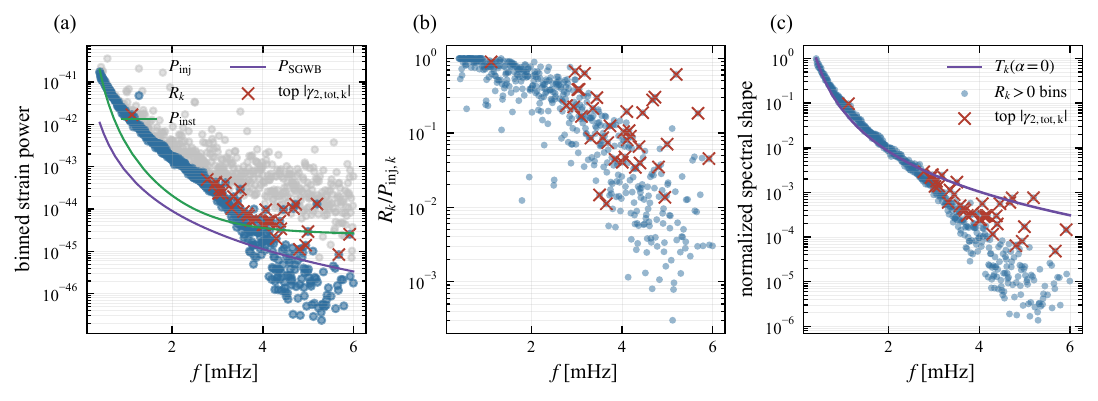}
    \caption{
Residual foreground power across the \(0.4\)--\(6.0\,{\rm mHz}\) analysis
band, evaluated with the orbit-averaged long-wavelength Michelson \(X\)
response and divided by \({\cal R}_{\rm LW}^{X}\) in
Eq.~\eqref{eq:lwa_x_response}. Of the
560 equal-width bins, 487 contain nonzero residual power. Panel (a) compares the
injected Galactic binary power \(P_{{\rm inj},k}\) (gray points), residual
power \(R_k\) (blue points), instrumental power
\(P_{{\rm inst},k}\) (green curve), and fiducial \(\alpha=0\),
\(\Omega_0=10^{-11}\) SGWB power \(P_{{\rm SGWB},k}\) (purple curve).
Panel (b) shows the residual fraction
\(R_k/P_{{\rm inj},k}\) as blue points. Panel (c) compares the
residual spectrum \(R_k\) (blue points) and the \(\alpha=0\) SGWB spectrum
\(T_k(0)\) (purple curve), each normalized by its maximum over the analysis
band. The red crosses mark the
\(N_{\rm bin}=36\) subset with the largest values of
\(|\gamma_{2,\rm tot,k}|\), defined in Sec.~\ref{sec:covariance}.
}
    \label{fig:all_bin_residual_power_distribution}
\end{figure*}

Figure~\ref{fig:all_bin_residual_power_distribution} compares the binned
injected and residual Galactic binary powers with the instrumental noise and the fiducial
\(\alpha=0\), \(\Omega_0=10^{-11}\) SGWB contribution. In panel (a), gray
points show \(P_{{\rm inj},k}\), while blue points show nonzero
\(R_k\). Because the residual sources are a subset of the
injections, \(R_k\leq P_{{\rm inj},k}\); the two points coincide
when no source in the bin meets the recovery criteria. The red crosses
identify the \(N_{\rm bin}=36\) subset with
the largest values of \(|\gamma_{2,\rm tot,k}|\), defined in
Sec.~\ref{sec:covariance}.

The residual fraction is largest in parts of the low-frequency foreground
where many binaries occupy the same catalog bin and the recovery criteria
leave a comparatively large fraction of the injected power in the residual.
The residual spectrum therefore retains appreciable low-frequency power and
can partially align with the \(\alpha=0\) SGWB strain spectrum, whose
Fourier-frequency dependence is \(f^{-3}\), even though
\(\Omega_{\rm GW}\) is frequency independent. This partial spectral alignment
contributes to the inverse-variance-weighted residual--SGWB degeneracy
quantified in Sec.~\ref{sec:model}.

\subsection{Residual source discreteness}
\label{sec:concentration}

Consider one frequency bin and either its injected or residual source
population. Using the source powers \(p_i\) in
Eq.~\eqref{eq:response_normalized_source_power}, we introduce the
normalized weights
\begin{equation}
    w_i
    =
    \frac{p_i}{\sum_{\ell} p_{\ell}}\, ,
\end{equation}
where the sum is over the specified source population in that bin. We quantify
source discreteness by the dimensionless quantity
\begin{equation}
    C_2
    =
    \sum_i w_i^2\, .
    \label{eq:c2}
\end{equation}
This quadratic concentration measure of the normalized source powers
corresponds to the ``source power concentration'' in the title.
It measures how the bin power
is shared among the contributing sources, rather than their number per
frequency interval. It is small when many sources contribute comparable
power, and it approaches one when a single source dominates the bin.
For residual sources in a bin \(k\) with \(R_k>0\),
\begin{equation}
    C_{2,k}^{\rm res}
    =
    \frac{S_{2,k}^{\rm res}}{R_k^2}\, .
\end{equation}

To characterize the finite-source departure from Gaussian statistics, we
consider a sum \(x_{\rm res}=\sum_i x_i\) for the residual sources
in one frequency bin, with \(x_i=\sqrt{2p_i}\cos\phi_i\). Here \(p_i\) is the
orbit-averaged source power of
Eq.~\eqref{eq:response_normalized_source_power}, and the phases \(\phi_i\)
are independent and uniformly distributed on \([0,2\pi)\). Holding these
powers fixed while randomizing the phases isolates how the sharing of power
among a finite number of sources affects the excess kurtosis of the sum.
Appendix~\ref{app:random_phase_moments} derives the moment relations. Within
this construction, the source discreteness determines the residual excess
kurtosis:
\begin{equation}
    \gamma_{2,\rm res}
    =
    -\frac{3}{2}
    \frac{\sum_i p_i^2}
         {\left(\sum_{\ell} p_{\ell}\right)^2}
    =
    -\frac{3}{2}C_2\, .
    \label{eq:gamma2_c2}
\end{equation}
Thus \(C_2\ll1\) corresponds to a residual sum formed by many comparably
weighted sources, whereas \(C_2=1\) is the single-source limit. Within this
model, the residual source sum is platykurtic relative to a Gaussian
distribution. Equation~\eqref{eq:gamma2_c2} is exact for this sum,
which is not the orbit-modulated Michelson time
series. The excess kurtosis characterizes source discreteness through a
normalized fourth-order cumulant of the sum. It does not by itself determine
the covariance of binned Fourier power, which is developed in
Sec.~\ref{sec:covariance}.

A Monte Carlo validation using \(5\times10^4\) random phase realizations
per bin gives, across the 487 bins with \(R_k>0\), a Pearson correlation
coefficient \(r=0.9992\) between the analytic and Monte Carlo excess
kurtoses, with a root-mean-square difference
\(\operatorname{rms}(\gamma_{2,\rm MC}+3C_2/2)=1.70\times10^{-2}\).

\section{Covariance of binned Fourier power and source discreteness}
\label{sec:covariance}

Source discreteness enters the analysis in two distinct ways. For the sum
defined in Sec.~\ref{sec:concentration}, the excess kurtosis is determined
by the catalog source discreteness. The
covariance relevant for SGWB inference, however, is the covariance of
binned Fourier power and must be constructed from the Fourier-mode
statistics below.

The LDC2A observation contains \(N=6\,307\,200\) samples at a cadence of
\(5\,{\rm s}\), giving
\begin{equation}
    T_{\rm obs}=31\,536\,000\,{\rm s}\, ,
    \qquad
    \delta f_{\rm F}=T_{\rm obs}^{-1}
    =3.17098\times10^{-8}\,{\rm Hz}\, .
    \label{eq:fourier_spacing}
\end{equation}
The positive, non-DC, non-Nyquist Fourier frequencies are
\(f_j=j\delta f_{\rm F}\). Let \({\cal J}_k\) be the set of these frequencies
in bin \(k\), and let
\(M_k=|{\cal J}_k|\). The ratio of the analysis-bin width to the Fourier
spacing is \(\Delta f/\delta f_{\rm F}=\Delta fT_{\rm obs}=315.36\). Since
\(M_k\) counts discrete frequencies, it is either 315 or 316; these integer
counts enter the binned mean and covariance below.

Let \(\widehat S_j\) denote the one-sided power in the Fourier mode at \(f_j\),
expressed as a spectral density. For statistically independent complex
Gaussian Fourier amplitudes of a stationary process, the mode powers are
exponentially distributed and satisfy
\begin{equation}
    \left\langle\widehat S_j\right\rangle=S_j\, ,
    \qquad
    {\rm Cov}(\widehat S_j,\widehat S_{j'})
    =S_j^2\delta_{jj'}\, .
    \label{eq:fourier_power_moments}
\end{equation}
The binned power is
\begin{equation}
    \widehat V_k
    =
    \delta f_{\rm F}\sum_{j\in{\cal J}_k}\widehat S_j\, .
    \label{eq:binned_power_definition}
\end{equation}
Its mean and covariance are consequently
\begin{align}
    \mu_k
    &=
    \delta f_{\rm F}\sum_{j\in{\cal J}_k}S_j\, ,
    \\
    {\rm Cov}(\widehat V_k,\widehat V_{k'})
    &=
    \delta_{kk'}{\rm Var}_{\rm G}(\widehat V_k)\, ,
    \\
    {\rm Var}_{\rm G}(\widehat V_k)
    &=
    (\delta f_{\rm F})^2
    \sum_{j\in{\cal J}_k}S_j^2\, .
    \label{eq:fourier_bandpower_covariance}
\end{align}
\({\rm Var}_{\rm G}(\widehat V_k)\) denotes the binned-power variance under
this Gaussian Fourier-amplitude model and is determined by the
Fourier-mode spectral densities in bin \(k\).

Let \({\cal T}_j(\alpha)\) denote the one-sided SGWB strain spectral-density
template per unit \(\Omega_0\) at Fourier frequency \(f_j\); its explicit
form is given in Sec.~\ref{sec:model}. The total spectral density at
\(f_j\) in bin \(k\) is then
\begin{equation}
    S_j
    =
    S_{\rm inst}(f_j)
    +\Omega_0{\cal T}_j(\alpha)
    +\beta S_{{\rm res},j}\, .
    \label{eq:mode_spectrum}
\end{equation}
At the fiducial parameters, define the Gaussian-component spectral density
\(G_j\) and the residual spectral density \(H_j\) as
\begin{equation}
    G_j
    \equiv
    S_{\rm inst}(f_j)
    +\Omega_{0,\rm fid}{\cal T}_j(\alpha)\, ,
    \qquad
    H_j
    \equiv
    \beta_{\rm fid}S_{{\rm res},j}\, .
    \label{eq:stochastic_and_residual_mode_power}
\end{equation}
The Gaussian Fourier-amplitude covariance in
Eq.~\eqref{eq:fourier_power_moments} assigns the mode-power variance
\((G_j+H_j)^2\). The \(H_j^2\) term is the contribution to the mode-power
variance that arises from treating the residual Fourier amplitude as
Gaussian. For a finite-source treatment,
let \(H_{ij}\) be the contribution
of residual source \(i\) to \(H_j\), so that \(H_j=\sum_iH_{ij}\), and quantify
source discreteness within each Fourier mode with \(H_j>0\) by
\begin{equation}
    c_j
    =
    \frac{\sum_iH_{ij}^2}{H_j^2}\, .
    \label{eq:mode_source_concentration}
\end{equation}
When every source power is distributed uniformly among all Fourier frequencies
in its analysis bin, \(c_j=C_{2,k}^{\rm res}\) for
\(j\in{\cal J}_k\).

For comparison with the Gaussian Fourier-amplitude model, we consider a
finite-source random-phase (RP) model. Each residual Fourier
amplitude is modeled as a finite sum of source contributions with fixed
source powers and independent uniform phases, using the phase assumption
of Sec.~\ref{sec:concentration}. The combined instrumental-noise and SGWB
amplitude is a zero-mean complex Gaussian variable independent of those
phases. We calculate the mean and variance of \(\widehat S_j\), the mode
power expressed as a spectral density, by averaging over both the source
phases and this Gaussian component.

We use a diagonal covariance matrix for the mode powers
\(\widehat S_j\), retaining their variances and neglecting covariance
between distinct Fourier frequencies, whether within the same analysis
bin or in different bins. Multiplying the retained variances by \((\delta f_{\rm F})^2\)
and summing within each analysis bin gives the binned-power variance
\({\rm Var}_{\rm RP}(\widehat V_k)\). It differs from
Eq.~\eqref{eq:fourier_bandpower_covariance} because the residual amplitude
is a generally non-Gaussian sum of finitely many source contributions
rather than a Gaussian variable with the same mean power. The variance of the power in one Fourier
mode is derived in Appendix~\ref{app:random_phase_moments}.

In the fixed-residual (FR) case, the
fiducial residual complex amplitude is held fixed in each mode with power
\(\delta f_{\rm F}H_j\). Only the independent Gaussian instrumental and
SGWB component, with mean mode power \(\delta f_{\rm F}G_j\), is averaged
over. Summing the resulting diagonal mode-power variances gives
\({\rm Var}_{\rm FR}(\widehat V_k)\). For the uniform frequency
distribution, the three binned-power variances are
\begin{align}
    {\rm Var}_{\rm G}(\widehat V_k)
    &=
    (\delta f_{\rm F})^2
    \sum_{j\in{\cal J}_k}(G_j+H_j)^2\, ,
    \notag\\
    {\rm Var}_{\rm RP}(\widehat V_k)
    &=
    (\delta f_{\rm F})^2
    \sum_{j\in{\cal J}_k}
    \left[(G_j+H_j)^2-C_{2,k}^{\rm res}H_j^2\right]\, ,
    \notag\\
    {\rm Var}_{\rm FR}(\widehat V_k)
    &=
    (\delta f_{\rm F})^2
    \sum_{j\in{\cal J}_k}\left(G_j^2+2G_jH_j\right)\, .
    \label{eq:residual_covariance_treatments}
\end{align}
At the same fiducial spectral densities \(G_j\) and \(H_j\), the three
cases have the same mean binned power. The residual-only contributions to
\({\rm Var}(\widehat S_j)\) are \(H_j^2\), \((1-c_j)H_j^2\), and zero
in the Gaussian, RP, and FR cases, respectively. For RP, source-phase
averaging gives the reduction \(c_jH_j^2\) relative to the Gaussian result;
fixing the residual amplitude
removes its own power fluctuations entirely. Expanding the first two expressions in
Eq.~\eqref{eq:residual_covariance_treatments} shows that all three
contain the same \(2G_jH_j\) contribution from the fluctuating
cross term between the Gaussian component and the residual amplitude.
For the uniform distribution,
\(c_j=C_{2,k}^{\rm res}\), so relative to the Gaussian result the RP
binned variance is reduced by
\((\delta f_{\rm F})^2C_{2,k}^{\rm res}
\sum_{j\in{\cal J}_k}H_j^2\), whereas fixing the
residual amplitudes removes the full term
\((\delta f_{\rm F})^2\sum_{j\in{\cal J}_k}H_j^2\). Thus
\({\rm Var}_{\rm FR}\leq{\rm Var}_{\rm RP}\leq{\rm Var}_{\rm G}\) for these
diagonal variance models.

The Gaussian model does not retain finite-source phase structure.
RP includes its contribution to the diagonal variance, but omits
cross-frequency covariance generated by source phases shared across modes.
FR evaluates the covariance conditional on fixed fiducial residual
amplitudes and does not average over their variability; this does not
require fixing \(\beta\) in the mean model. All three omit
cross-frequency covariance, source-parameter uncertainty, spectral leakage
associated with the finite observation window, and the time-dependent
detector response. Their variance ordering does not by itself
determine the strength of the residual--SGWB parameter degeneracy or
establish bounds on the full cross-frequency covariance of a coherent LISA
residual.

The binned residual model retains the integrated catalog power \(R_k\) but does
not specify individual binary lines on the one-year Fourier grid. In the
uniform distribution, \(R_k\) is spread over the Fourier frequencies in its
analysis bin:
\begin{equation}
    S_{{\rm res},j}
    =
    \frac{R_k}{M_k\delta f_{\rm F}}\, ,
    \qquad j\in{\cal J}_k\, ,
    \label{eq:residual_psd_within_bin}
\end{equation}
which satisfies
\(\delta f_{\rm F}\sum_{j\in{\cal J}_k}S_{{\rm res},j}=R_k\).
For fixed \(R_k\), this distribution minimizes
\(\sum_{j\in{\cal J}_k}(\delta f_{\rm F}S_{{\rm res},j})^2\) and therefore
maximizes the effective number of residual Fourier modes in the bin. We assess
the dependence on unresolved frequency structure with two more concentrated
distributions. One distributes each source power equally among the Fourier frequencies spanning its
one-year intrinsic frequency drift, with the interval extended by the annual
Doppler half-width. The other assigns each source power to the closest
Fourier frequency. All three distributions conserve each source's total power. The
finite-source diagonal random-phase covariance is evaluated for the uniform
distribution, for which \(c_j=C_{2,k}^{\rm res}\). The alternative frequency
distributions are compared using the Gaussian Fourier-amplitude and
fixed-residual covariances with the same Fourier-power assignments.
These assignments distribute source powers without calculating the
coherent Fourier response over the observation window. The resulting
diagonal models do not include coherent sidebands, spectral leakage,
or cross-frequency correlations.
Define
\begin{equation}
    P_{{\rm inst},k}
    =\delta f_{\rm F}\sum_{j\in{\cal J}_k}S_{\rm inst}(f_j)\, ,
    \qquad
    T_k(\alpha)
    =\delta f_{\rm F}\sum_{j\in{\cal J}_k}{\cal T}_j(\alpha)\, ,
    \label{eq:mode_integrated_templates}
\end{equation}
so that the mean binned power is
\begin{equation}
    \mu_k
    =P_{{\rm inst},k}+\Omega_0T_k(\alpha)+\beta R_k\, .
    \label{eq:power_model}
\end{equation}
For each alternative residual distribution, \(R_k\) in
Eq.~\eqref{eq:power_model} is recomputed by summing the powers assigned to the
Fourier frequencies in bin \(k\). The same powers also determine
Eq.~\eqref{eq:fourier_bandpower_covariance},
so transfer between bins enters both the mean and the covariance.
For each specified spectral index \(\alpha\), we write
\(\mu_{k,\rm fid}\) for this mean at
\(\Omega_{0,\rm fid}=10^{-11}\) and \(\beta_{\rm fid}=1\). The covariance in
Eq.~\eqref{eq:fourier_bandpower_covariance} is evaluated at the same fiducial
parameters and held fixed in the Fisher information matrix.
When \(\Omega_{0,\rm fid}\) is varied in Sec.~\ref{sec:all_bin_results}, the
mode powers and covariance are recomputed at each amplitude before being held
fixed in the corresponding local Fisher information matrix.

Under the independent Gaussian Fourier-amplitude model, the binned power
\(\widehat V_k\) is a sum of \(M_k\) independent exponential variables
\(\delta f_{\rm F}\widehat S_j\), whose mean values are generally unequal.
We match its mean and variance to those of
\((\mu_{k,\rm fid}/\nu_{{\rm eff},k})\chi^2_{\nu_{{\rm eff},k}}\),
where \(\chi^2_\nu\) denotes a chi-square variable with \(\nu\) degrees of
freedom. This defines the effective number of degrees of freedom
\(\nu_{{\rm eff},k}=2\mu_{k,\rm fid}^2/{\rm Var}_{\rm G}(\widehat V_k)\).
This representation is exact when all modes in the bin have equal
mean power, giving \(\nu_{{\rm eff},k}=2M_k\). Unequal mean mode
powers give \(\nu_{{\rm eff},k}<2M_k\), and higher moments need not match.

For the uniform residual frequency distribution at the fiducial parameters
with \(\alpha=0\), each analysis bin contains 315 or 316 modes with nearly
equal mean powers. The effective number of degrees of freedom remains close
to \(2M_k\), with a minimum of approximately 630. Direct evaluation of the
third and fourth cumulants of the sum of independent exponential mode powers
gives skewness approximately \(0.11251\)--\(0.11275\) and excess kurtosis
approximately \(0.01899\)--\(0.01907\) across all 560 bins. The large number
of independent contributions and the small skewness and excess kurtosis
support the Gaussian approximation to the binned power in this baseline.
The minimum effective number of degrees of freedom is approximately 93 when
each source power is spread over its intrinsic drift
and Doppler frequency range, and 3.3 when it is assigned to the nearest
Fourier frequency. Concentrating power among fewer
Fourier modes reduces the effective number of degrees of freedom. The
Gaussian approximation is least well justified in the latter case.

To quantify how instrumental and SGWB power dilute the excess kurtosis
of the sum defined in Sec.~\ref{sec:concentration}, let
\(\gamma_{2,\rm res,k}\) and \(\gamma_{2,\rm tot,k}\) denote the excess
kurtoses of the residual and total scalar fluctuations in bin \(k\).
At the fiducial parameters, adding the independent Gaussian instrumental
and SGWB contributions gives
\begin{equation}
    \gamma_{2,\rm tot,k}
    =
    \gamma_{2,\rm res,k}
    \left(
    \frac{\beta_{\rm fid}R_k}{\mu_{k,\rm fid}}
    \right)^2\, .
    \label{eq:gamma2_dilution}
\end{equation}

\begin{figure*}[t]
    \centering
    \includegraphics[
        width=0.95\textwidth
    ]{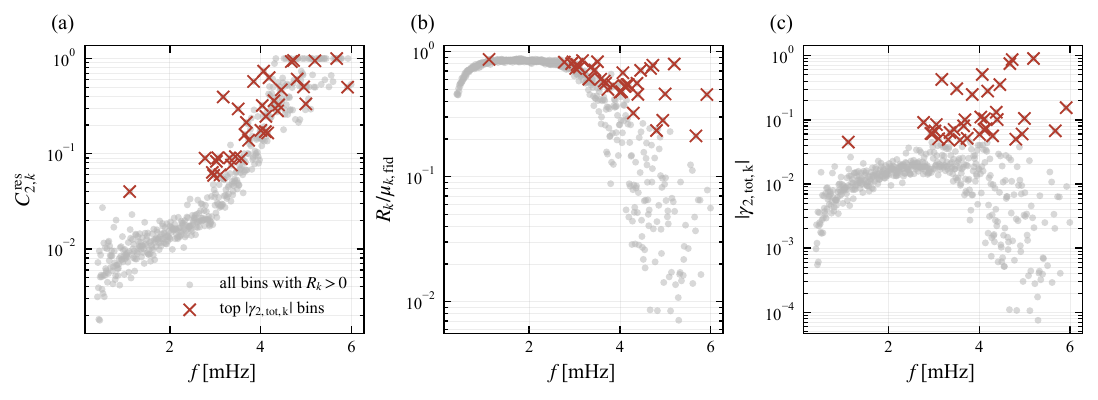}
    \caption{
Source discreteness and dilution by instrumental and SGWB power in the 487 bins with \(R_k>0\), using
the orbit-averaged long-wavelength Michelson \(X\) source powers. Panel (a)
shows \(C_{2,k}^{\rm res}\). Panel (b) shows the residual fraction of the
fiducial total power, \(R_k/\mu_{k,\rm fid}\). Panel (c) shows the magnitude
of the diluted excess kurtosis, \(|\gamma_{2,\rm tot,k}|\), defined in
Eq.~\eqref{eq:gamma2_dilution}. The red crosses mark the
\(N_{\rm bin}=36\) bins with the largest values of
\(|\gamma_{2,\rm tot,k}|\). Panels (b) and (c) use
\(\Omega_0=10^{-11}\), \(\alpha=0\), and \(\beta_{\rm fid}=1\).
}
    \label{fig:all_bin_noise_dilution}
\end{figure*}

The cumulative subsets used in Sec.~\ref{sec:gamma_prior_results} rank the 487
bins with \(R_k>0\) by \(|\gamma_{2,\rm tot,k}|\), evaluated at
\(\Omega_0=10^{-11}\), \(\alpha=0\), and \(\beta_{\rm fid}=1\). The
\(N_{\rm bin}=6,12,24,36\) subsets contain the bins with the largest values of
\(|\gamma_{2,\rm tot,k}|\) and are held fixed when the three SGWB slopes are compared.
Fig.~\ref{fig:all_bin_noise_dilution} shows the source discreteness, residual power
fraction, and diluted excess kurtosis across the analysis band.

For \(\beta_{\rm fid}=1\) used in Fig.~\ref{fig:all_bin_noise_dilution},
Eqs.~\eqref{eq:gamma2_c2} and \eqref{eq:gamma2_dilution} link the three
panels: the magnitude in panel (c) is proportional to the source discreteness
in panel (a) times the squared residual fraction in panel (b).
Around \(2\,{\rm mHz}\), the residual contributes most of the fiducial
total power, but its power is shared among many sources, giving small source
discreteness and a small excess kurtosis in magnitude. At higher frequencies,
source discreteness increases while the residual fraction generally decreases,
producing stronger dilution. The larger values of
\(|\gamma_{2,\rm tot,k}|\) occur in bins where increased source discreteness
coincides with a residual contribution that remains an appreciable fraction
of the total power.

The median \(|\gamma_{2,\rm tot,k}|\) over the 487 bins is approximately
\(1.4\times10^{-2}\), although a small set of bins reaches much larger values,
with a maximum of approximately \(0.90\).
In this model, \(\gamma_{2,\rm tot,k}\) is negative
because the residual sum is platykurtic.

Determining the full cross-frequency non-Gaussian covariance of the binned
powers would require the joint fourth-order correlations of the Fourier
amplitudes, including spectral leakage and the time-dependent detector
response. Those correlations are not
fixed by \(C_2\) or by Eq.~\eqref{eq:gamma2_dilution}. The Gaussian
Fourier-amplitude covariance in
Eq.~\eqref{eq:fourier_bandpower_covariance} is therefore used as the
baseline covariance for the SGWB inference, with the finite-source
random-phase and fixed-residual diagonal treatments providing the
comparisons defined above. The frequency subsets are ranked by
\(|\gamma_{2,\rm tot,k}|\), whereas the prior-width requirement for \(\beta\) and the
omitted-residual shift depend on the Fisher elements constructed from
\(R_k\) and \(T_k(\alpha)\), including their inverse-variance-weighted
spectral overlap.

\section{Binned SGWB model and residual-power uncertainty}
\label{sec:model}

We quantify the spectral degeneracy between an isotropic SGWB and the
catalog residual in the mean binned power. The two parameters are the SGWB
amplitude \(\Omega_0\) and the dimensionless factor \(\beta\) multiplying the
residual power. We derive their Fisher information matrix and distinguish the
uncertainty increase caused by marginalizing over \(\beta\) from the bias
caused by omitting the residual component.

\subsection{Binned power model and Fisher information}
\label{sec:binned_power_fisher}

Equation~\eqref{eq:power_model} gives the mean power in bin \(k\).
Here \(P_{{\rm inst},k}\), \(T_k(\alpha)\), and \(R_k\) are the binned
instrumental power, SGWB strain-power template per unit \(\Omega_0\), and
catalog residual power, respectively. The
fiducial value is \(\beta_{\rm fid}=1\).

We model the dimensionless SGWB energy density as
\begin{equation}
    \Omega_{\rm GW}(f)
    =
    \Omega_0
    \left(
    \frac{f}{f_{\rm ref}}
    \right)^\alpha\, ,
\end{equation}
where \(f_{\rm ref}=3\,{\rm mHz}\) is the reference frequency and \(\alpha\)
is held fixed in each Fisher calculation. The corresponding one-sided strain power spectral
density is~\cite{Maggiore2000SGWB}
\begin{equation}
    S_h(f)
    =
    \frac{3H_0^2}{2\pi^2}
    \frac{\Omega_{\rm GW}(f)}{f^3}\, ,
\end{equation}
where we set \(H_0=67.4\,{\rm km\,s^{-1}\,Mpc^{-1}}\)
~\cite{2020A&A...641A...6P}. The spectral density per
unit \(\Omega_0\) at Fourier frequency \(f_j\) is
\begin{equation}
    {\cal T}_j(\alpha)
    =
    \frac{3H_0^2}{2\pi^2}
    \frac{(f_j/f_{\rm ref})^\alpha}{f_j^3}\, .
    \label{eq:sgwb_mode_template}
\end{equation}
The binned template \(T_k(\alpha)\) is its Fourier-frequency sum in
Eq.~\eqref{eq:mode_integrated_templates}, and
\(P_{{\rm SGWB},k}=\Omega_0T_k(\alpha)\).
Because \(S_h(f)\) has units of strain squared per hertz,
\(P_{{\rm SGWB},k}\), \(P_{{\rm inst},k}\), and \(R_k\) all have units of
strain squared. Both \(\Omega_0\) and \(\beta\) are dimensionless.

Unless stated otherwise, we evaluate the Gaussian Fourier-amplitude covariance
in Eq.~\eqref{eq:fourier_bandpower_covariance} at the fiducial parameters and
hold it fixed, so the parameter dependence enters only through the mean
spectrum.
In the baseline two-parameter analysis, the instrumental noise spectrum is
treated as known and is not included among the inferred parameters.

The quantities \(\gamma_{2,\rm res,k}\) and \(\gamma_{2,\rm tot,k}\)
describe the excess kurtosis of a sum of source signals with independent
random phases, before and after adding the independent Gaussian instrumental
and SGWB components, respectively. These are distinct from the excess
kurtosis of the binned power \(\widehat V_k\). Under the independent Gaussian
Fourier-amplitude model, the large number of contributing modes and the
small skewness and excess kurtosis calculated in Sec.~\ref{sec:covariance}
support the Gaussian approximation for the uniform baseline. The variance ordering in
Eq.~\eqref{eq:residual_covariance_treatments} compares second moments only
and does not establish Gaussianity.

For \(\bm{\theta}=(\Omega_0,\beta)\), we approximate the binned-power
likelihood as Gaussian, with a covariance that is diagonal in frequency bins
and fixed at its fiducial value. The Fisher information matrix is
\begin{equation}
    F_{ab}
    =
    \sum_k
    \frac{
    \partial_{\theta_a}\mu_k
    \partial_{\theta_b}\mu_k
    }
    {{\rm Var}_{\rm G}(\widehat V_k)}\, .
    \label{eq:fisher}
\end{equation}
The finite-source random-phase and fixed-residual comparisons retain
the same local Gaussian approximation to the binned-power likelihood.
\({\rm Var}_{\rm G}(\widehat V_k)\) in Eq.~\eqref{eq:fisher} is replaced by
\({\rm Var}_{\rm RP}(\widehat V_k)\) and
\({\rm Var}_{\rm FR}(\widehat V_k)\), respectively, from
Eq.~\eqref{eq:residual_covariance_treatments}; the mean and its parameter
derivatives are unchanged.
Here \(a\) and \(b\) index the parameters \(\Omega_0\) and \(\beta\).
The Fisher information matrix describes the expected local curvature of the
negative log likelihood at the fiducial point
~\cite{CutlerFlanagan1994ParameterEstimation,Vallisneri2008FisherMatrix}. The
derivatives are
\(\partial_{\Omega_0}\mu_k=T_k(\alpha)\) and
\(\partial_\beta\mu_k=R_k\). Accordingly, the elements are denoted
\(F_{\Omega\Omega}\), \(F_{\Omega\beta}\), and \(F_{\beta\beta}\). The
off-diagonal element \(F_{\Omega\beta}\) measures the overlap between the
changes in binned power produced by \(\Omega_0\) and \(\beta\), after weighting
each bin by its inverse variance.
The alternative variances change these frequency weights and hence the
weighted spectral overlap and amplitude uncertainties. These comparisons
do not evaluate the exact Fisher information of the corresponding full non-Gaussian
distributions.

If \(\beta\) is fixed, the uncertainty on \(\Omega_0\) is
\(\sigma_{\Omega,\rm fixed}=F_{\Omega\Omega}^{-1/2}\). If \(\beta\) is
inferred jointly, the marginalized uncertainty is
\(\sigma_{\Omega,\rm marg}=[(F^{-1})_{\Omega\Omega}]^{1/2}\). We characterize
the local precision by the corresponding fractional uncertainties,
\(\sigma_{\Omega,\rm fixed}/\Omega_{0,\rm fid}\) and
\(\sigma_{\Omega,\rm marg}/\Omega_{0,\rm fid}\). These quantities describe
the local uncertainty on the SGWB amplitude within the adopted
fixed-covariance model. A
Gaussian prior \(\beta\sim{\cal N}(1,\sigma_\beta^2)\)
adds \(1/\sigma_\beta^2\) to \(F_{\beta\beta}\):
\begin{equation}
    F_{\beta\beta}
    \rightarrow
    F_{\beta\beta}+\frac{1}{\sigma_\beta^2}\, .
\end{equation}
The inverse of the Fisher matrix including this prior contribution gives the
posterior covariance implied by the adopted fixed-covariance Gaussian
likelihood and prior.

If the fiducial residual is present in the mean power but omitted from the
fitted model, the resulting shift in the best-fitting SGWB amplitude is
\begin{equation}
    \Delta\Omega_{\rm bias}
    =
    \beta_{\rm fid}
    \frac{F_{\Omega\beta}}{F_{\Omega\Omega}}\, ,
    \qquad
    \frac{\Delta\Omega_{\rm bias}}{\sigma_{\Omega,\rm fixed}}
    =
    \beta_{\rm fid}
    \frac{F_{\Omega\beta}}{\sqrt{F_{\Omega\Omega}}}\, .
    \label{eq:linearized_bias}
\end{equation}
For the linear mean model in Eq.~\eqref{eq:power_model}, with
\(\alpha\), the instrumental noise spectrum, and the covariance
fixed at their fiducial values, only \(\Omega_0\) is fitted when
the residual is omitted. The shift \(\Delta\Omega_{\rm bias}\)
is the inverse-variance-weighted projection of the omitted residual
spectrum onto the SGWB template. We do not calculate a bias in \(\alpha\)
in the present analysis. If \(\alpha\) were also inferred,
the omitted residual would generally shift both fitted parameters.
Within the same fixed-covariance approximation, their first-order
shifts would be obtained using the Fisher matrix for
\((\Omega_0,\alpha)\) and the weighted projections of the omitted
residual onto the corresponding derivatives of the mean spectrum.
The numerical results use \(\beta_{\rm fid}=1\). This shift is
distinct from the uncertainty increase produced when \(\beta\)
is included and marginalized.

In the limit of no prior on \(\beta\), define the uncertainty increase factor
\begin{equation}
    I_0
    \equiv
    \frac{\sigma_{\Omega,\rm marg}}
         {\sigma_{\Omega,\rm fixed}}
    =
    \frac{1}{\sqrt{1-\rho_{\Omega\beta}^2}}\, ,
    \qquad
    \rho_{\Omega\beta}
    \equiv
    \frac{F_{\Omega\beta}}
         {\sqrt{F_{\Omega\Omega}F_{\beta\beta}}}\, .
    \label{eq:uncertainty_increase}
\end{equation}
\(\rho_{\Omega\beta}\) is the normalized off-diagonal Fisher
element and therefore an inverse-variance-weighted spectral overlap. For the
two-parameter Fisher matrix without a prior, the corresponding correlation
coefficient from the parameter covariance matrix is
\(-\rho_{\Omega\beta}\). Consequently, only \(|\rho_{\Omega\beta}|\) enters
\(I_0\).

\subsection{Prior width for a specified uncertainty threshold}
\label{sec:prior_requirement}

For a specified SGWB slope \(\alpha\), set of frequency bins, and
allowed increase in the \(\Omega_0\) uncertainty, we ask how weak the
Gaussian prior on the residual-power factor \(\beta\) can be while
keeping the uncertainty penalty from jointly fitting \(\beta\) within
the chosen threshold. Equivalently, we determine the largest prior
standard deviation \(\sigma_\beta\) satisfying
\begin{equation}
    \frac{\sigma_{\Omega,\rm marg}(\sigma_\beta)}
         {\sigma_{\Omega,\rm fixed}}
    \leq
    r_{\rm max}\, .
    \label{eq:prior_requirement}
\end{equation}
Here \(r_{\rm max}\) is the maximum allowed ratio of the marginalized
uncertainty to the fixed-\(\beta\) uncertainty. The values
\(r_{\rm max}=1.1,1.2,1.5,2.0\) limit the increase in the
\(\Omega_0\) uncertainty to \(10\%\), \(20\%\), \(50\%\), and \(100\%\),
respectively. Appendix~\ref{app:fisher_prior} gives the analytic solution for
the corresponding prior width.

The value \(I_0\) without a prior is the
\(\sigma_\beta\rightarrow\infty\) limit of the left side of
Eq.~\eqref{eq:prior_requirement}. If \(I_0\leq r_{\rm max}\), the condition is
already satisfied with \(\beta\) freely marginalized and places no finite upper
bound on \(\sigma_\beta\). Such cases are denoted by \(\infty\) in
Table~\ref{tab:prior_requirement}. In the opposite limit,
\(\sigma_\beta\rightarrow0\), the prior fixes \(\beta\) and the ratio
approaches unity. Within the fixed-covariance two-parameter model, the ratio
increases monotonically as the prior is weakened.
Section~\ref{sec:finite_bin_results} applies
these definitions to all 560 bins and to frequency subsets.

\section{Residual--SGWB degeneracy and prior widths}
\label{sec:finite_bin_results}

The full-band result uses all 560 bins. Subsets of the 487 bins with nonzero
residual power show how restricting the frequency set changes the spectral
degeneracy, the allowed prior width for \(\beta\), and the omitted-residual
bias. Unless another covariance is specified, the results use the Gaussian
Fourier-amplitude covariance. Here
\(N_{\rm bin}\) denotes the number of bins in a subset. We denote by
\(\sigma_\beta^{x\%}\) the largest
prior width \(\sigma_\beta\) satisfying
Eq.~\eqref{eq:prior_requirement} with
\(r_{\rm max}=1+x/100\). Thus, for example,
\(\sigma_\beta^{10\%}\) is the largest prior width that limits the increase in
the marginalized uncertainty on \(\Omega_0\) to \(10\%\).

\subsection{All-bin results and SGWB-amplitude dependence}
\label{sec:all_bin_results}

For \(\alpha=0\), \(\Omega_{\rm GW}\) is independent of frequency. With
Gaussian residual Fourier amplitudes and the residual power distributed
uniformly among the Fourier frequencies in each bin, the Fisher information
matrix using all 560 bins gives
\(|\rho_{\Omega\beta}|\simeq0.475\) and \(I_0\simeq1.1361\). The
fixed-\(\beta\) fractional uncertainty is
\(\sigma_{\Omega,\rm fixed}/\Omega_{0,\rm fid}\simeq2.09\%\).
After fitting \(\beta\) freely, it becomes
\(\sigma_{\Omega,\rm marg}/\Omega_{0,\rm fid}\simeq2.37\%\).
Table~\ref{tab:all_bin_covariances} summarizes these all-bin results.
Marginalizing freely over \(\beta\) increases the \(\Omega_0\) uncertainty by
\(13.6\%\) with Gaussian residual amplitudes. The finite-source diagonal
random-phase covariance changes this increase to \(13.8\%\), a difference of
\(0.2\) percentage points. Although the random-phase covariance gives
slightly smaller absolute uncertainties, it also changes the
inverse-variance weighting and hence the normalized residual--SGWB
spectral overlap. In this uniform all-bin calculation, the
fixed-\(\beta\) uncertainty is reduced by a slightly larger fraction
than the marginalized uncertainty, so the relative marginalization
penalty is slightly larger. At a \(10\%\) threshold, the corresponding values
are \(\sigma_\beta^{10\%}\simeq0.00728\) and
\(\sigma_\beta^{10\%}\simeq0.00698\), respectively.

\begin{table}[t]
\centering
\caption{
All-bin results for \(\alpha=0\), \(\Omega_{0,\rm fid}=10^{-11}\), and the
uniform residual frequency distribution. The percentage increase is
\(100(I_0-1)\), and \(\sigma_\beta^{10\%}\) is the largest prior width that
limits this increase to \(10\%\).
}
\label{tab:all_bin_covariances}
\begin{ruledtabular}
\begin{tabular}{lccc}
Residual covariance & \(I_0\) & Increase & \(\sigma_\beta^{10\%}\) \\
\hline
Gaussian amplitudes & 1.1361 & \(13.6\%\) & 0.00728 \\
Finite-source random phase & 1.1383 & \(13.8\%\) & 0.00698 \\
\end{tabular}
\end{ruledtabular}
\end{table}

The 73 bins with \(R_k=0\) contribute to \(F_{\Omega\Omega}\), but not to
\(F_{\Omega\beta}\) or \(F_{\beta\beta}\). They therefore improve the
fixed-\(\beta\) constraint on \(\Omega_0\) without adding residual--SGWB
spectral overlap. The result restricted to the 487 bins with \(R_k>0\) is
given in Appendix~\ref{app:additional_checks}.

Figure~\ref{fig:sgwb_amplitude_dependence} shows the dependence on the fiducial
SGWB amplitude for \(10^{-13}\leq\Omega_0\leq10^{-10}\). At every point, the
mean mode powers and covariance are recomputed at that amplitude and the
resulting covariance is then held fixed in the local Fisher information matrix.
The source catalog and its source-discreteness values are held fixed;
the two curves compare the Gaussian Fourier-amplitude covariance
with the finite-source random-phase correction for this population.
For Gaussian residual amplitudes, the uncertainty increase rises from
\(12.7\%\) at \(10^{-13}\) to \(20.7\%\) at \(10^{-10}\); the finite-source
diagonal calculation gives \(12.9\%\) and \(20.8\%\), respectively. The close
agreement between the two covariances therefore extends across this
amplitude range.

The fixed-\(\beta\) fractional uncertainty decreases from approximately
\(176\%\) at \(10^{-13}\) to \(0.469\%\) at \(10^{-10}\), passing through
\(2.09\%\) at the fiducial value. The corresponding marginalized
fractional uncertainty is this quantity multiplied by \(I_0\), and equals
\(2.37\%\) at \(\Omega_0=10^{-11}\). At
\(\Omega_0=10^{-13}\), even the fixed-\(\beta\) uncertainty exceeds the
fiducial amplitude, so the local Fisher result does not describe a precise
amplitude measurement. The prior width in
Fig.~\ref{fig:sgwb_amplitude_dependence}(b) is not monotonic because changing
\(\Omega_0\) changes the inverse-variance weights in all three Fisher elements.
Unlike \(I_0\), which depends on the normalized spectral overlap,
the largest allowed prior width also depends on the absolute
Fisher information. In this calculation, the finite-source
correction has a larger fractional effect on the allowed prior
width than on \(I_0\) or the fixed-\(\beta\) fractional uncertainty,
and a slightly tighter prior is needed to meet the \(10\%\) threshold.

\begin{figure*}[t]
    \centering
    \includegraphics[
        width=0.95\textwidth
    ]{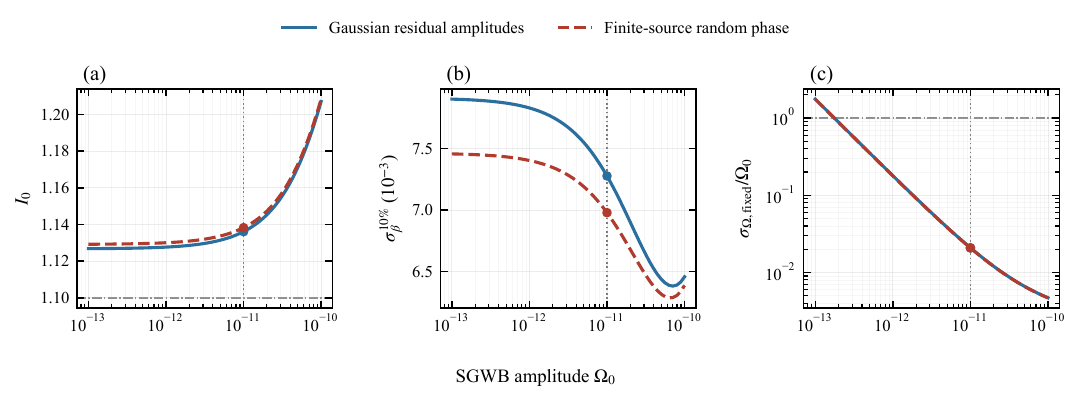}
    \caption{
All-bin dependence on the fiducial SGWB amplitude for \(\alpha=0\),
\(\beta_{\rm fid}=1\), and residual power distributed uniformly among the
Fourier frequencies in each analysis bin. At each \(\Omega_0\), the covariance
is evaluated at that fiducial amplitude and then held fixed in the local Fisher
information matrix. Panel (a) shows the uncertainty increase factor \(I_0\)
without a prior on \(\beta\); the horizontal line marks \(I_0=1.1\). Panel (b) shows the largest
prior width \(\sigma_\beta^{10\%}\) that limits the increase to \(10\%\).
Panel (c) shows the fixed-\(\beta\) fractional uncertainty,
\(\sigma_{\Omega,\rm fixed}/\Omega_0\);
the horizontal line marks unity (\(100\%\) fractional uncertainty). The
vertical dotted line and circles mark \(\Omega_0=10^{-11}\). Solid and dashed
curves use the Gaussian Fourier-amplitude and finite-source diagonal random-phase
covariances, respectively. The corresponding fractional uncertainty after
fitting \(\beta\) is the quantity in panel (c) multiplied by \(I_0\) in panel (a).
}
    \label{fig:sgwb_amplitude_dependence}
\end{figure*}

Omitting the fiducial residual from the same 560-bin model gives
\(\Delta\Omega_{\rm bias}\simeq2.49\times10^{-11}\), with
\(\Delta\Omega_{\rm bias}/\sigma_{\Omega,\rm fixed}\simeq119.5\). This is the
inverse-variance-weighted projection of the omitted residual spectrum onto the
SGWB template in Eq.~\eqref{eq:linearized_bias}, with the covariance held fixed;
it is not a posterior detection significance.

The preceding all-bin results use the uniform residual frequency
distribution. Distributing each source power among the Fourier frequencies
covered by its one-year intrinsic drift and annual Doppler range instead gives
\(|\rho_{\Omega\beta}|\simeq0.438\), \(I_0\simeq1.1124\), and
\(\sigma_\beta^{10\%}\simeq0.01451\), corresponding to an uncertainty increase
of \(11.2\%\). For comparison, conditioning the covariance on the fiducial
residual Fourier realization gives a \(21.0\%\) increase for the uniform
frequency distribution while retaining the free factor \(\beta\) in the mean.
Additional results for assigning each source power to the nearest Fourier
frequency, the
fixed-residual covariance, and a free instrumental-noise amplitude are given
in Appendix~\ref{app:additional_checks}. The uncertainty increase therefore depends
on how the residual power is distributed among Fourier frequencies and on the
adopted covariance, while the Gaussian and finite-source diagonal covariances
give nearly identical increases for the all-bin uniform calculation.

\subsection{Subsets ranked by the magnitude of the diluted excess kurtosis}
\label{sec:gamma_prior_results}

The following subset results use the Gaussian Fourier-amplitude covariance
with residual power distributed uniformly among the Fourier frequencies in
each analysis bin.
The subsets ranked by \(|\gamma_{2,\rm tot,k}|\) contain the bins in which the
magnitude of the source-sum excess kurtosis remains largest after adding instrumental and
fiducial SGWB power. The ranking is based on the diluted source-sum excess
kurtosis, whereas the corresponding prior widths are computed from the
Fisher matrix. For \(\alpha=0\), the
\(N_{\rm bin}=6,12,24,36\) subsets give \(I_0\) between \(1.76\) and \(3.04\)
and \(\sigma_\beta^{10\%}\) between \(0.0087\) and \(0.0149\).

The larger \(I_0\) values reflect stronger residual--SGWB spectral
correlation in these subsets. At the same time, the subsets contain less
total Fisher information, so their allowed prior widths can remain broader
than the all-bin value despite the larger marginalization penalty.

These subset uncertainties follow from the local Fisher approximation. For the
smallest subsets,
\(\sigma_{\Omega,\rm fixed}\) can be of order \(\Omega_{0,\rm fid}\) or larger.
These values should therefore be interpreted as local Gaussian Fisher
uncertainties rather than posterior credible intervals; the physical
boundary \(\Omega_0\geq0\) lies outside this approximation.

Table~\ref{tab:prior_requirement} in
Appendix~\ref{app:additional_checks} reports the subset results for all three
SGWB spectral slopes. Because \(\Omega_{0,\rm fid}=10^{-11}\) enters the
covariance weights, the tabulated values are conditional on this fiducial
amplitude and change for another choice. The fiducial amplitude serves as
the expansion point of the local Fisher calculation rather than as a LISA
sensitivity threshold.

\begin{figure*}[t]
    \centering
    \includegraphics[
        width=0.95\textwidth
    ]{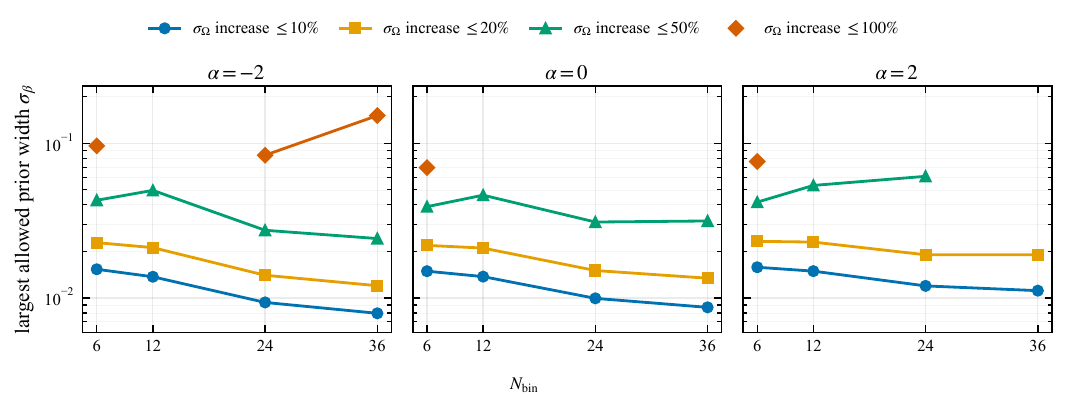}
    \caption{
Prior standard deviations for \(\beta\) in subsets ranked by
\(|\gamma_{2,\rm tot,k}|\). The three
panels correspond to SGWB spectral slopes \(\alpha=-2,0,2\), with
\(\Omega_0=10^{-11}\). For each \(N_{\rm bin}=6,12,24,36\), the subset
contains the bins with the largest \(|\gamma_{2,\rm tot,k}|\) among the 487
bins with \(R_k>0\).
Each curve gives the largest \(\sigma_\beta\) that limits the marginalized
\(\Omega_0\) uncertainty increase to the indicated level. Only bounded values
are plotted; cases that meet the threshold with \(\beta\) freely marginalized
are omitted. For the thresholds tabulated in Table~\ref{tab:prior_requirement},
the corresponding entries are listed as \(\infty\). The residual power in each
analysis bin is distributed uniformly among its one-year Fourier frequencies.
The source powers are evaluated with the long-wavelength Michelson \(X\)
response averaged over orbital phase, and the covariance is given by
Eq.~\eqref{eq:fourier_bandpower_covariance}.
}
    \label{fig:prior_requirement}
\end{figure*}

Figure~\ref{fig:prior_requirement} shows the prior widths for the
\(10\%\), \(20\%\), \(50\%\), and \(100\%\) thresholds as functions of
\(N_{\rm bin}\). In the \(\alpha=0\) panel,
\(\sigma_\beta^{10\%}\) decreases from \(0.0149\) at \(N_{\rm bin}=6\) to
\(0.0087\) at \(N_{\rm bin}=36\). Over the same subset sizes, \(I_0\) changes
from \(3.04\) to \(1.76\), with a local increase between
\(N_{\rm bin}=12\) and 24. Adding bins changes both the
inverse-variance-weighted spectral overlap between \(T_k\) and \(R_k\) and
the total Fisher information that sets the numerical prior scale. The
\(\alpha=2\) spectrum shows weaker residual--SGWB spectral degeneracy
for the largest subset. For the \(50\%\) threshold, the prior-width bound
is finite for \(N_{\rm bin}=6,12,24\), whereas the \(N_{\rm bin}=36\) subset
already satisfies the threshold with \(\beta\) freely marginalized.

Only bounded values of \(\sigma_\beta\) are plotted in
Fig.~\ref{fig:prior_requirement}. When \(I_0\leq r_{\rm max}\), the
corresponding point is omitted. For the \(10\%\), \(20\%\), and \(50\%\)
thresholds tabulated in Table~\ref{tab:prior_requirement}, the corresponding
unbounded cases appear as \(\infty\). The largest allowed prior width rises
sharply as \(I_0\) approaches \(r_{\rm max}\) from above. The prior-width
bound need not vary monotonically with \(N_{\rm bin}\), because the three
Fisher elements
\(F_{\Omega\Omega}\), \(F_{\Omega\beta}\), and \(F_{\beta\beta}\) accumulate
at different rates. Adding a bin can therefore increase or decrease
\(|\rho_{\Omega\beta}|\), and hence \(I_0\), even though the subset grows
cumulatively.

Alternative bin widths and frequency subset selections are examined in
Appendix~\ref{app:additional_checks}. At \(\alpha=0\), for
\(N_{\rm bin}=12,24,36\), rankings by \(|\gamma_{2,\rm tot,k}|\) and by the
per-bin cross-Fisher magnitude
\(\left|T_k(0)R_k/{\rm Var}_{\rm G}(\widehat V_k)\right|\) give comparable
prior widths, whereas ranking by
\(T_k^2(0)/{\rm Var}_{\rm G}(\widehat V_k)\) gives substantially broader values.
The subset results therefore depend on the selected frequencies and on all
three Fisher elements; the all-bin result is independent of the
ranking rule.

\section{Discussion and conclusions}
\label{sec:discussion}

This work quantifies how residual-power uncertainty and residual--SGWB
spectral overlap affect SGWB inference at the catalog level. Because the residual and SGWB spectra overlap, uncertainty in the residual power weakens the SGWB amplitude constraint when the two amplitudes are inferred jointly, while omitting the residual biases the fitted SGWB amplitude. Separately, we characterize the
statistics of the summed residual source signals using independent random
phases and amplitudes determined by their orbit-averaged powers.

This catalog-level analysis complements previous LISA studies of the
Galactic-foreground problem.
Buscicchio~\emph{et~al.}~\cite{Buscicchio2025ForegroundGaussianity} examined
the Gaussianity and stationarity of the time-dependent Galactic white-dwarf
foreground, while Rosati and
Littenberg~\cite{Rosati2024SGWBResidual} studied SGWB recovery directly from
global-fit residuals and quantified the impact of residual non-Gaussianity.

For \(\alpha=0\), \(\Omega_0=10^{-11}\), and all 560 bins with residual power
distributed uniformly over the Fourier frequencies within each bin,
\(|\rho_{\Omega\beta}|\simeq0.475\) and
\(I_0\simeq1.136\). Marginalizing freely over \(\beta\) increases the
uncertainty on \(\Omega_0\) by \(13.6\%\) with the Gaussian Fourier-amplitude
covariance and by \(13.8\%\) with the finite-source diagonal random-phase
covariance. Under the Gaussian Fourier-amplitude covariance, the largest
prior width that limits the uncertainty increase to \(10\%\) is
\(\sigma_\beta^{10\%}\simeq0.0073\).

Within this random phase construction, the source discreteness
\(C_2=\sum_i w_i^2\) determines the residual excess kurtosis,
\(\gamma_{2,\rm res}=-3C_2/2\). Instrumental and fiducial SGWB power reduce its
magnitude to a median \(|\gamma_{2,\rm tot,k}|\simeq1.4\times10^{-2}\) over
the 487 bins with nonzero residual power, although a small number of bins remain
strongly platykurtic. The binned Fourier-power covariance additionally depends
on joint fourth-order correlations among Fourier amplitudes.

Marginalizing over \(\beta\) increases the all-bin uncertainty throughout
\(10^{-13}\leq\Omega_0\leq10^{-10}\). With Gaussian residual amplitudes, the
increase changes from \(12.7\%\) at \(\Omega_0=10^{-13}\) to \(20.7\%\) at
\(10^{-10}\); the maximum difference in the uncertainty increase between the
Gaussian Fourier-amplitude and random-phase covariances is \(0.24\)
percentage points over this range. The
dependence arises because \(\Omega_0\) enters the fiducial mode powers and
therefore changes the inverse-variance weights, even though the mean-spectrum
derivatives remain \(T_k\) and \(R_k\).

The uncertainty increase and the largest allowed prior width also depend on
the Fourier-frequency distribution of the residual power and the covariance. With Gaussian residual
Fourier amplitudes, spreading each source power over its intrinsic drift
and Doppler frequency range gives an \(11.2\%\) uncertainty increase. Conditioning the covariance on a fixed residual Fourier
realization gives a \(21.0\%\) increase for the uniform distribution. In both
cases the residual--SGWB spectral degeneracy persists; the associated
uncertainty increase and prior-width requirement vary across these modeling
assumptions. For the uniform distribution and Gaussian residual amplitudes,
omitting the fiducial residual gives
\(\Delta\Omega_{\rm bias}\simeq2.49\times10^{-11}\), with
\(\Delta\Omega_{\rm bias}/\sigma_{\Omega,\rm fixed}\simeq119.5\). The shift
\(\Delta\Omega_{\rm bias}\) is the inverse-variance-weighted projection of the
residual spectrum onto the SGWB template with the covariance held fixed.
Its ratio to \(\sigma_{\Omega,\rm fixed}\) is not a posterior detection
significance. The subset results further show that the allowed prior
width, uncertainty increase, and omitted-residual shift have different
dependences on the selected frequencies and the three Fisher elements.

The Fisher uncertainties considered here characterize local parameter precision
rather than the detection significance of an SGWB in observed LISA data. A
realistic channel-level detection analysis would additionally model
multichannel time-delay interferometry (TDI) data together with instrumental
noise and the Galactic
foreground~\cite{AdamsCornish2014SGWBForeground,Boileau2021SpectralSeparation};
it would also incorporate uncertainty in the instrumental noise
model~\cite{Muratore2024NoiseUncertainty}. Under their respective detector and
foreground models, Adams and Cornish recovered a scale-invariant background
down to \(2\times10^{-13}\) with six links and \(5\times10^{-13}\) with four
links, while Boileau \emph{et al.} reported detectability near
\(8\times10^{-13}\) using the \(A,E,T\) channels. Because those studies use
different detector responses, foreground models, noise treatments, and
inference frameworks, their detection thresholds are not directly comparable to
the fractional uncertainties used here. They nevertheless place
\(\Omega_0=10^{-11}\) above the background amplitudes recovered in those
simulations. The present calculation quantifies how the catalog residual
changes the local uncertainty at this fiducial amplitude.

The analysis is defined at the catalog level. The residual spectrum \(R_k\) is
constructed from injected sources that do not meet the adopted recovery
criteria, and its uncertainty is represented by the single normalization
parameter \(\beta\). Source powers are evaluated with an orbit-averaged
long-wavelength Michelson \(X\) response and expressed as equivalent incident
strain. This construction isolates uncertainty in the overall residual power
and its spectral degeneracy with the SGWB. A channel-level analysis would
additionally account for posterior uncertainty in recovered binaries, imperfect
subtraction, residual spectral-shape uncertainty, finite-arm and
time-dependent response effects, coherent sidebands, and cross-channel
correlations. For observed LISA data, the analogous residual spectrum would be
inferred from global-fit posterior residuals or an uncertain foreground model.

The covariance is diagonal in frequency bins and fixed at the fiducial
spectrum. In the Gaussian Fourier-amplitude model, the mode-power variance
includes the \(H_j^2\) contribution from independent Gaussian residual Fourier
amplitudes. The finite-source diagonal random-phase covariance is obtained
by averaging over independent source phases for the uniform frequency
distribution, whereas the fixed-residual covariance conditions on one Fourier
realization. The two-parameter Fisher
information matrix fixes the instrumental noise spectrum; the three-parameter
extension allows its overall amplitude to vary. The Fisher uncertainties follow from a local Gaussian approximation to
binned power. For the uniform Gaussian Fourier-amplitude baseline, this
approximation is supported by the many independent mode powers and the
small skewness and excess kurtosis calculated in
Sec.~\ref{sec:covariance}. For more concentrated residual-power
assignments within the Gaussian Fourier-amplitude model, the effective
number of contributing Fourier modes is smaller, so the Gaussian
approximation is less well supported. In the uniform baseline, using the random-phase
covariance instead of the Gaussian covariance changes the marginalization
penalty only slightly. These comparisons do not determine how departures
from Gaussianity or neglected cross-frequency correlations affect parameter
inference. The numerical
prior width depends on the frequency range, bin width, residual frequency
distribution, instrumental-noise model, and residual covariance.

Further catalog-level studies can vary the Galactic binary population,
recovery criteria, and residual spectral model. A channel-level extension
could propagate global-fit residual realizations through the time-dependent
LISA response and evaluate the resulting cross-frequency and cross-channel
covariance. These extensions would quantify the impact of population uncertainty
and assess how the time-dependent detector response to the Galactic
foreground and multichannel information may improve residual--SGWB
separation. Across the comparisons made here, the residual--SGWB spectral degeneracy
remains a persistent contribution to the SGWB uncertainty. For the fiducial
Gaussian Fourier-amplitude baseline, keeping the uncertainty increase at or below
\(10\%\) requires \(\sigma_\beta\lesssim0.0073\), motivating the propagation
of residual foreground uncertainty from global fits into future LISA SGWB
analyses.

\begin{acknowledgments}
This work was supported by the National Key Research and Development Program
of China under Grant Nos. 2023YFC2206700 and 2022YFC2205201, the National Natural Science Foundation of China (NSFC) under Grant No. 12673046, and by the
Major Science and Technology Program of Xinjiang Uygur Autonomous Region under
Grant No. 2022A03013-4. Computations were performed on the High Performance
Computing Platform of Huazhong University of Science and Technology. We thank the anonymous referee for useful suggestions that improved the manuscript.
\end{acknowledgments}

\section*{Data Availability}

The processed data and analysis code that support the findings of
this article are openly available~\cite{GuanWang2026Code}.
The input catalogs are available from the LDC2A data
set~\cite{Baghi2022LDCZonedo} and the Erebor LDC2A training
catalogs~\cite{Katz2024EreborLDC2A}.

\appendix

\section{Additional results for covariance, binning, and frequency subsets}
\label{app:additional_checks}

\subsection{All-bin comparisons}

The 73 bins with \(R_k=0\) constrain \(\Omega_0\) without contributing to
\(F_{\Omega\beta}\) or \(F_{\beta\beta}\). Restricting the calculation with the
Gaussian Fourier-amplitude covariance to the 487 bins with \(R_k>0\)
gives \(I_0\simeq1.1799\) and
\(\sigma_\beta^{10\%}\simeq0.00503\), compared with \(1.1361\) and
\(0.00728\) when all 560 bins are included. The 73 zero-residual bins therefore
reduce the normalized spectral overlap by adding information on \(\Omega_0\)
without adding residual power.

For the uniform residual frequency distribution, the finite-source diagonal
random-phase covariance gives a marginalized uncertainty of
\(2.363\times10^{-13}\), compared with \(2.371\times10^{-13}\) for Gaussian
residual amplitudes. Thus the Gaussian Fourier-amplitude covariance gives a slightly
smaller relative increase, \(13.6\%\) rather than \(13.8\%\), but a slightly
larger absolute marginalized uncertainty. The smaller absolute marginalized
uncertainty under the random-phase covariance reflects its smaller overall
variance, whereas the relative increase is slightly larger because the
fixed-\(\beta\) uncertainty is reduced by a slightly larger fraction
than the marginalized uncertainty.

Assigning each source power to the closest Fourier frequency gives
\(|\rho_{\Omega\beta}|\simeq0.308\) and \(I_0\simeq1.0510\) with Gaussian
residual amplitudes, so the \(10\%\) threshold is satisfied without an
informative prior. We use this assignment only to examine how
concentrating each source's power in one Fourier frequency changes the result;
it is not a coherent binary-line model. The three frequency distributions
therefore show the dependence on unresolved frequency structure rather than
define physical bounds on a coherent LISA signal.

Conditioning the covariance on the fixed fiducial residual Fourier realization
changes the variance while retaining the mean model in
Eq.~\eqref{eq:power_model} and its free factor \(\beta\). For the uniform
distribution, this covariance gives
\(|\rho_{\Omega\beta}|\simeq0.563\), \(I_0\simeq1.2102\), and an uncertainty
increase of \(21.0\%\). The corresponding largest prior width is
\(\sigma_\beta^{10\%}\simeq0.00258\). Spreading each source power over its
intrinsic drift and Doppler frequency range gives an increase of \(21.1\%\),
while assigning it to the nearest Fourier frequency gives \(21.0\%\), with
all three prior widths approximately \(0.00258\). The decrease from \(13.6\%\)
to \(5.1\%\) found with Gaussian residual amplitudes is therefore not present
when the residual Fourier realization is fixed.

Equation~\eqref{eq:residual_covariance_treatments} explains this difference.
For these frequency distributions, the term \(\sum_j2G_jH_j\) depends mainly
on the total residual power in an analysis bin because \(G_j\) varies slowly
across that bin. The Gaussian Fourier-amplitude covariance also contains
\(\sum_jH_j^2\), which grows when a fixed total residual power is concentrated
into fewer Fourier frequencies and reduces their inverse-variance weight.
Frequency concentration therefore changes the result obtained with the
Gaussian Fourier-amplitude covariance mainly through the \(H_j^2\) variance term.
With the uniform distribution, the
fixed-residual covariance gives
\(\Delta\Omega_{\rm bias}\simeq4.47\times10^{-11}\) and
\(\Delta\Omega_{\rm bias}/\sigma_{\Omega,\rm fixed}\simeq239.7\). This is the
inverse-variance-weighted projection of the omitted residual spectrum onto
the SGWB template, with the covariance held fixed; it is not a posterior
detection significance.

We also allow the overall instrumental-noise amplitude to vary under the
Gaussian Fourier-amplitude model by setting
\(\mu_k=\eta P_{{\rm inst},k}+\Omega_0T_k(\alpha)+\beta R_k\), with
\(\eta_{\rm fid}=1\). The covariance remains fixed at its fiducial value in
this three-parameter Fisher matrix. Marginalizing freely over \(\eta\), while
keeping the frequency dependence of the instrumental spectrum fixed, gives
\(I_0\simeq1.1299\) and \(\sigma_\beta^{10\%}\simeq0.00848\). With \(\beta\)
fixed, marginalizing over \(\eta\) increases the fractional uncertainty on
\(\Omega_0\) from \(2.09\%\) to \(5.88\%\). Fitting \(\beta\) as well
gives a fractional uncertainty of \(6.65\%\). In this
definition of \(I_0\), \(\eta\) is marginalized in both uncertainties;
\(\beta\) is fixed in the denominator and marginalized in the numerator.
Uncertainty in the overall instrumental-noise amplitude therefore has a larger
effect on the absolute SGWB constraint than on the relative increase caused by
marginalizing over \(\beta\). This three-parameter extension isolates uncertainty in the overall
instrumental-noise normalization; a more general noise model would also
allow uncertainty in its spectral shape.

When the residual power is distributed uniformly among the Fourier frequencies
in each bin, the bin width changes both the binned mean and covariance. For
\(\alpha=0\), the all-bin values are \(I_0\simeq1.119\), \(1.136\), and
\(1.157\) for bin widths of \(5\), \(10\), and \(20\,\mu{\rm Hz}\),
respectively. The corresponding \(10\%\) prior widths are approximately
\(0.0101\), \(0.00728\), and \(0.00583\). The band-integrated residual,
instrumental, and fiducial SGWB powers are unchanged to numerical precision;
the differences arise from the frequency grouping of the component spectra and
their mode-sum variances.

\subsection{Frequency subset comparisons}

Table~\ref{tab:prior_requirement} gives the numerical results for the subsets
ranked by \(|\gamma_{2,\rm tot,k}|\) and discussed in
Sec.~\ref{sec:gamma_prior_results}.

\begin{table*}[t]
\centering
\caption{
Prior standard deviations for \(\beta\) in subsets selected from the
487 bins with
nonzero residual power, evaluated at \(\Omega_{0,\rm fid}=10^{-11}\) and
the covariance in Eq.~\eqref{eq:fourier_bandpower_covariance},
with the residual spectral density constant among the Fourier frequencies in
each analysis bin.
Each subset contains the \(N_{\rm bin}\) bins with the largest values of
\(|\gamma_{2,\rm tot,k}|\). The fixed-\(\beta\) fractional uncertainty
\(\sigma_{\Omega,\rm fixed}/\Omega_{0,\rm fid}\) is defined in
Sec.~\ref{sec:model}. \(I_0\) is the
uncertainty increase factor without a prior on \(\beta\),
\(|\rho_{\Omega\beta}|\) is the magnitude of the normalized off-diagonal
Fisher element for \(\Omega_0\) and \(\beta\), and
\(\Delta\Omega_{\rm bias}/\sigma_{\Omega,\rm fixed}\) is the shift
for \(\beta_{\rm fid}=1\) when the residual foreground is omitted, as defined in
Eq.~\eqref{eq:linearized_bias}. In \(\sigma_\beta^{x\%}\), \(x\%\) is the
allowed fractional increase in the
marginalized \(\Omega_0\) uncertainty. Entries \(\infty\) indicate that the
specified threshold is met without an informative prior on \(\beta\).
}
\label{tab:prior_requirement}
\begin{ruledtabular}
\begin{tabular}{ccccccccc}
\(\alpha\) &
\(N_{\rm bin}\) &
\(\sigma_{\Omega,\rm fixed}/\Omega_{0,\rm fid}\) &
\(I_0\) &
\(|\rho_{\Omega\beta}|\) &
\(\Delta\Omega_{\rm bias}/\sigma_{\Omega,\rm fixed}\) &
\(\sigma_\beta^{10\%}\) &
\(\sigma_\beta^{20\%}\) &
\(\sigma_\beta^{50\%}\)
\\
\hline
\(-2\) & 6  & 0.784  & 2.36 & 0.906 & 30.60 & 0.015 & 0.023 & 0.043 \\
\(-2\) & 12 & 0.374  & 1.76 & 0.824 & 35.22 & 0.014 & 0.021 & 0.050 \\
\(-2\) & 24 & 0.213  & 2.15 & 0.885 & 50.47 & 0.009 & 0.014 & 0.027 \\
\(-2\) & 36 & 0.143  & 2.03 & 0.870 & 59.64 & 0.008 & 0.012 & 0.024 \\
\(0\)  & 6  & 0.437  & 3.04 & 0.944 & 31.16 & 0.015 & 0.022 & 0.039 \\
\(0\)  & 12 & 0.211  & 1.84 & 0.840 & 34.90 & 0.014 & 0.021 & 0.046 \\
\(0\)  & 24 & 0.131  & 1.97 & 0.862 & 47.86 & 0.010 & 0.015 & 0.031 \\
\(0\)  & 36 & 0.100  & 1.76 & 0.824 & 55.54 & 0.009 & 0.013 & 0.031 \\
\(2\)  & 6  & 0.221  & 2.90 & 0.939 & 29.42 & 0.016 & 0.023 & 0.042 \\
\(2\)  & 12 & 0.112  & 1.77 & 0.826 & 32.36 & 0.015 & 0.023 & 0.053 \\
\(2\)  & 24 & 0.072  & 1.60 & 0.780 & 41.17 & 0.012 & 0.019 & 0.061 \\
\(2\)  & 36 & 0.057  & 1.42 & 0.712 & 46.08 & 0.011 & 0.019 & \(\infty\) \\
\end{tabular}
\end{ruledtabular}
\end{table*}

These subset results show that the numerical prior width is set jointly by
the normalized spectral overlap and the absolute Fisher information carried
by the selected bins.

\subsection{Dependence on frequency subset selection}

The \(|\gamma_{2,\rm tot,k}|\) ranking selects bins with the largest magnitude
of the excess kurtosis after dilution, whereas the prior width depends on all three
Fisher elements. To assess the dependence on the subset definition, we also
consider rankings by two per-bin Fisher contributions, approximately equally
spaced frequency selections, and random subsets. All selections use the same
487 bins with \(R_k>0\), with \(\Omega_0=10^{-11}\), \(\alpha=0\), and the
Gaussian Fourier-amplitude covariance.

The first additional ranking uses the magnitude of
\(F_{\Omega\beta,k}\equiv
T_k(0)R_k/{\rm Var}_{\rm G}(\widehat V_k)\), the per-bin contribution to the
off-diagonal Fisher element. The second uses
\(T_k^2(0)/{\rm Var}_{\rm G}(\widehat V_k)\), the contribution to
\(F_{\Omega\Omega}\), and therefore emphasizes bins that constrain
\(\Omega_0\) when \(\beta\) is fixed. We also select approximately equally
spaced frequency ranks and draw 200 random subsets without replacement at each
subset size. The frequency-rank and random selections are constructed
independently for each value of \(N_{\rm bin}\).

\begin{figure*}[t]
    \centering
    \includegraphics[
        width=0.95\textwidth
    ]{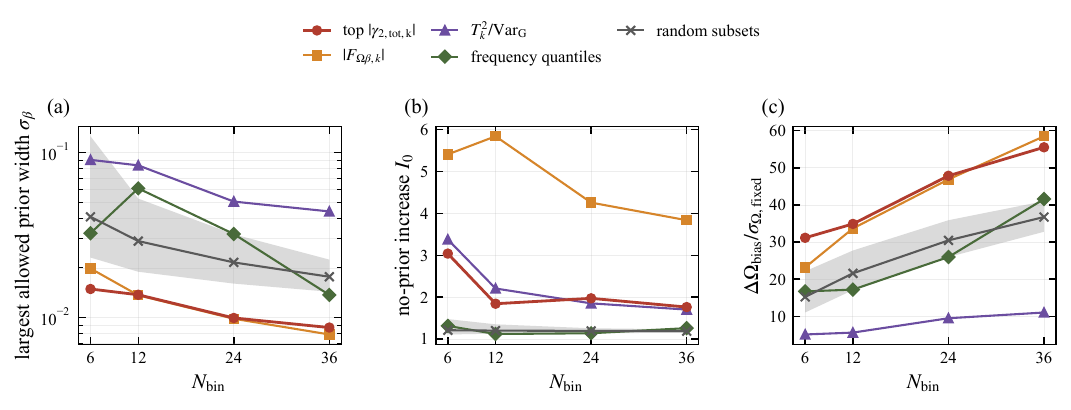}
    \caption{
Dependence on frequency subset selection for \(\Omega_0=10^{-11}\),
\(\alpha=0\), and a \(10\%\) threshold for the marginalized \(\Omega_0\)
uncertainty. Panel (a) shows the largest allowed prior width
\(\sigma_\beta^{10\%}\), panel (b) shows the no-prior uncertainty increase
factor \(I_0\), and panel (c) shows the normalized shift
\(\Delta\Omega_{\rm bias}/\sigma_{\Omega,\rm fixed}\) when the residual is
omitted. The gray crosses and bands show the median and 16th--84th percentiles
of the corresponding quantity over 200 random subsets at each size.
In panel (a), unbounded prior widths are retained as \(\infty\) when
computing these summary statistics over all 200 subsets. Only finite
summary values are plotted. The bands summarize
variation among frequency subsets rather than measurement confidence
intervals. The residual power is distributed uniformly among the Fourier
frequencies in each analysis bin.
}
    \label{fig:bin_selection_robustness}
\end{figure*}

Figure~\ref{fig:bin_selection_robustness}(a) shows that the numerical prior
width is most sensitive to the frequency selection for the smallest subsets.
The \(|\gamma_{2,\rm tot,k}|\) and \(|F_{\Omega\beta,k}|\) rankings give
similar prior widths for \(N_{\rm bin}=12,24,36\), whereas the
\(T_k^2/{\rm Var}_{\rm G}\) ranking gives substantially broader values. Thus
neither the diluted excess kurtosis nor the off-diagonal contribution alone
determines the prior width.

The \(|F_{\Omega\beta,k}|\) ranking gives the largest \(I_0\) at each displayed
subset size, while panel (c) follows a different pattern. In particular, the
normalized omitted-residual shift can grow while \(I_0\) decreases because
additional bins can improve the separation of the two spectral shapes while
increasing the residual projection relative to the fixed-\(\beta\) statistical
uncertainty. The marginalization penalty and the omitted-residual shift are
therefore distinct consequences of the same residual component.

\section{Random phase moments and finite-source diagonal covariance}
\label{app:random_phase_moments}

This appendix derives the moment relations connecting the dimensionless source
discreteness \(C_2\) to the excess kurtosis of the summed source signals,
its dilution by independent Gaussian components, and the
finite-source diagonal correction to the Fourier-power covariance.

Consider the residual source sum in an analysis bin,
\begin{equation}
    x_{\rm res}
    =
    \sum_i x_i\, ,
    \qquad
    x_i=\sqrt{2p_i}\cos\phi_i\, ,
\end{equation}
where the phases \(\phi_i\) are independent and uniformly distributed. Here
\(p_i\) is the binned strain power contribution defined in
Sec.~\ref{sec:data}, so
\(x_i\) has units of strain. This sum uses orbit-averaged source powers
and is not the time-dependent Michelson strain. For one source,
\begin{equation}
    \langle x_i\rangle=0\, ,
    \qquad
    \langle x_i^2\rangle=p_i\, ,
    \qquad
    \langle x_i^4\rangle=\frac{3}{2}p_i^2\, .
\end{equation}
The fourth cumulant of one source is therefore
\begin{equation}
    \kappa_{4,i}
    =
    \langle x_i^4\rangle
    -
    3\langle x_i^2\rangle^2
    =
    -\frac{3}{2}p_i^2\, .
\end{equation}
Since cumulants add for independent variables,
\begin{equation}
    \kappa_{4,\rm res}
    =
    -\frac{3}{2}\sum_i p_i^2\, ,
    \qquad
    \sigma_{\rm res}^2
    =
    \sum_i p_i\, .
\end{equation}
The residual excess kurtosis is therefore
\begin{equation}
    \gamma_{2,\rm res}
    =
    \frac{\kappa_{4,\rm res}}{\sigma_{\rm res}^4}
    =
    -\frac{3}{2}
    \frac{\sum_i p_i^2}{\left(\sum_{\ell} p_{\ell}\right)^2}
    =
    -\frac{3}{2}C_2\, .
\end{equation}
This is the relation used in Eq.~\eqref{eq:gamma2_c2}.

If the total fluctuation is
\begin{equation}
    x=x_{\rm res}+x_{\rm G}\, ,
\end{equation}
where \(x_{\rm G}\) is approximately Gaussian and statistically independent of
the residual component, define \(P_{\rm G}\equiv\langle x_{\rm G}^2\rangle\).
Under this Gaussian approximation, the fourth cumulant of \(x_{\rm G}\)
vanishes, so the fourth cumulant of the total field is
\(\kappa_{4,\rm res}\), while the total variance is
\begin{equation}
    \sigma_{\rm tot}^2
    =
    \sigma_{\rm res}^2+P_{\rm G}\, .
\end{equation}
The total excess kurtosis is therefore diluted to
\begin{equation}
    \gamma_{2,\rm tot}
    =
    \gamma_{2,\rm res}
    \left(
        \frac{\sigma_{\rm res}^2}{\sigma_{\rm tot}^2}
    \right)^2\, .
\end{equation}
Thus, within the independent random phase model, \(C_2\) determines the
source sum excess kurtosis, while instrumental and SGWB power dilute its
magnitude by the square of the residual variance fraction.

To derive the finite-source diagonal covariance used in Sec.~\ref{sec:covariance}, write
the residual complex amplitude at frequency \(f_j\) as
\begin{equation}
    s_j
    =
    \sum_i\sqrt{H_{ij}}e^{i(\phi_i+\psi_{ij})}\, ,
    \qquad
    H_j=\sum_iH_{ij}\, .
\end{equation}
Here \(\phi_i\) are independent uniform source phases and \(\psi_{ij}\) are
fixed mode-dependent phases. The source-phase average gives
\begin{equation}
    \left\langle|s_j|^2\right\rangle_\phi=H_j\, ,
    \qquad
    {\rm Var}_\phi\!\left(|s_j|^2\right)
    =H_j^2-\sum_iH_{ij}^2\, .
\end{equation}
Let \(g_j\) be an independent zero-mean complex Gaussian amplitude with
\(\langle|g_j|^2\rangle=G_j\). At fixed source phases,
\begin{equation}
    {\rm Var}_g\!\left(|g_j+s_j|^2\mid\{\phi_i\}\right)
    =G_j^2+2G_j|s_j|^2\, .
\end{equation}
Averaging this conditional variance over the source phases and adding the
variance of the conditional mean gives
\begin{align}
    {\rm Var}_{g,\phi}\!\left(|g_j+s_j|^2\right)
    &=
    (G_j+H_j)^2-\sum_iH_{ij}^2
    \notag\\
    &=
    (G_j+H_j)^2-c_jH_j^2\, ,
\end{align}
where \(c_j\) is defined in Eq.~\eqref{eq:mode_source_concentration}. If every
source power is distributed uniformly among the Fourier frequencies of its
analysis bin, \(H_{ij}/H_j=p_i/R_k\), and therefore
\(c_j=C_{2,k}^{\rm res}\). Summing these diagonal mode variances gives
\({\rm Var}_{\rm RP}\) in Eq.~\eqref{eq:residual_covariance_treatments}.
Because a given source phase is shared across Fourier frequencies, a full
cross-frequency covariance requires the mode-dependent amplitudes and
phases of a coherent response model. The calculation above provides the
finite-source diagonal variance used in Sec.~\ref{sec:covariance}.

\section{Marginalization and prior for the residual power factor}
\label{app:fisher_prior}

This appendix derives the analytic relations used to determine the largest
allowed prior width for \(\beta\) in the local two-parameter Fisher
information matrix with \(\bm{\theta}=(\Omega_0,\beta)\). A Gaussian prior
\(\beta\sim{\cal N}(1,\sigma_\beta^2)\) is imposed.

For a fixed spectral index \(\alpha\), with the argument of \(T_k(\alpha)\)
suppressed for brevity, the mean binned power is
\begin{equation}
    \mu_k
    =
    P_{{\rm inst},k}
    +
    \Omega_0 T_k
    +
    \beta R_k\, .
\end{equation}
Using the Gaussian covariance in
Eq.~\eqref{eq:fourier_bandpower_covariance}, fixed at the fiducial parameters,
we define
\begin{equation}
    A
    =
    F_{\Omega\Omega}
    =
    \sum_k
    \frac{T_k^2}{{\rm Var}_{\rm G}(\widehat V_k)}\, ,
\end{equation}
\begin{equation}
    B
    =
    F_{\beta\beta}
    =
    \sum_k
    \frac{R_k^2}{{\rm Var}_{\rm G}(\widehat V_k)}\, ,
\end{equation}
and
\begin{equation}
    C
    =
    F_{\Omega\beta}
    =
    \sum_k
    \frac{T_kR_k}{{\rm Var}_{\rm G}(\widehat V_k)}\, .
\end{equation}
The same relations give the finite-source diagonal random-phase or fixed-residual results
when \({\rm Var}_{\rm G}(\widehat V_k)\) is replaced throughout by
\({\rm Var}_{\rm RP}(\widehat V_k)\) or
\({\rm Var}_{\rm FR}(\widehat V_k)\), respectively.
If the residual-power factor \(\beta\) is held fixed, the fixed-\(\beta\)
SGWB amplitude uncertainty is
\begin{equation}
    \sigma_{\Omega,\rm fixed}
    =
    A^{-1/2}\, .
\end{equation}
A Gaussian prior \(\beta\sim{\cal N}(1,\sigma_\beta^2)\) replaces
\begin{equation}
    B
    \rightarrow
    B_\sigma
    =
    B+\frac{1}{\sigma_\beta^2}\, .
\end{equation}
After marginalizing over \(\beta\), the SGWB amplitude variance is
\begin{equation}
    \sigma_{\Omega,\rm marg}^2(\sigma_\beta)
    =
    \frac{B_\sigma}{AB_\sigma-C^2}\, .
\end{equation}
The uncertainty increase factor is therefore
\begin{equation}
    I^2(\sigma_\beta)
    =
    \left[
    \frac{\sigma_{\Omega,\rm marg}(\sigma_\beta)}
         {\sigma_{\Omega,\rm fixed}}
    \right]^2
    =
    \frac{AB_\sigma}{AB_\sigma-C^2}\, .
\end{equation}
In the limit with no prior, \(B_\sigma=B\), giving
\begin{equation}
    I_0
    =
    \frac{1}{\sqrt{1-\rho_{\Omega\beta}^2}}\, ,
    \qquad
    \rho_{\Omega\beta}
    =
    \frac{C}{\sqrt{AB}}\, .
\end{equation}
Here \(\rho_{\Omega\beta}\) is the normalized off-diagonal Fisher element. In
this two-parameter Fisher matrix without a prior, the corresponding correlation
coefficient obtained from the inverse Fisher information matrix is
\(-\rho_{\Omega\beta}\). The two quantities therefore have the same magnitude.

For a specified increase limit \(r_{\rm max}\), the largest allowed prior width,
corresponding to the weakest prior that still satisfies the limit, can be
written analytically. We impose
\begin{equation}
    I(\sigma_\beta)\leq r_{\rm max}\, .
\end{equation}
This condition is equivalent to
\begin{equation}
    B+\frac{1}{\sigma_\beta^2}
    \geq
    \frac{C^2}
    {A\left(1-r_{\rm max}^{-2}\right)}\, .
\end{equation}
Therefore, if
\begin{equation}
    B
    \geq
    \frac{C^2}
    {A\left(1-r_{\rm max}^{-2}\right)}\, ,
\end{equation}
the threshold is already satisfied without an informative prior, so
\(\sigma_{\beta,\max}=\infty\). Otherwise, the largest allowed prior width,
denoted by \(\sigma_{\beta,\max}\), is
\begin{equation}
    \sigma_{\beta,\max}
    =
    \left[
    \frac{C^2}
    {A\left(1-r_{\rm max}^{-2}\right)}
    -
    B
    \right]^{-1/2}\, .
\end{equation}
The prior width therefore depends jointly on the off-diagonal element
\(F_{\Omega\beta}=\sum_k T_kR_k/{\rm Var}_{\rm G}(\widehat V_k)\) and the two
diagonal elements \(F_{\Omega\Omega}\) and \(F_{\beta\beta}\).

\section{Catalog matching, detector response, and covariance of Fourier powers}
\label{app:numerical_conventions}

This appendix specifies the catalog association, orbit-averaged
long-wavelength response, instrumental noise, and Fourier-power covariance that
enter the binned SGWB inference.

\subsection{Frequency bins and recovered source labels}

We use the injected Galactic binary populations in the HDF5 catalog groups
\texttt{sky/dgb/cat} and \texttt{sky/igb/cat} of the LDC2A data
set~\cite{Baghi2022LDCZonedo}. Sources with
\(0.4\,{\rm mHz}\leq f_i<6.0\,{\rm mHz}\) are assigned to the unique
\(10\,\mu{\rm Hz}\) bin satisfying
\begin{equation}
    f_{k,{\rm left}}\leq f_i<f_{k,{\rm right}}\, .
\end{equation}
Recovery labels are obtained from the Erebor LDC2A injection-recovery
match~\cite{Katz2024EreborLDC2A}. A match row is accepted as recovered when
\begin{equation}
    C_{{\rm rec},i}\geq0.9\, ,
    \qquad
    {\cal O}_{{\rm best},i}\geq0.9\, .
\end{equation}
Each accepted match row is associated with a unique LDC2A source using
the eight injection-parameter fields shared by the two catalog products:
frequency, amplitude, frequency derivative, initial phase, inclination,
polarization, and the two ecliptic coordinates. Sources satisfying both
recovery criteria are classified as recovered and excluded from the catalog
residual; the remaining injected sources define the residual population.

\subsection{Orbit-averaged Michelson response and catalog sums}

We evaluate the source power with the long-wavelength Michelson \(X\) response
~\cite{Cutler1998AngularResolution,CornishRubbo2003LISAResponse}. At orbital
phase \(\varphi\), the first-order equal-arm LDC orbit specifies the relative
spacecraft geometry, up to a common displacement and overall scale, through
the dimensionless vectors \(\bm r_n(\varphi)\). With
\(\eta_n=2\pi n/3\),
\begin{equation}
    \bm r_n(\varphi)
    =
    \begin{pmatrix}
    \sin\varphi\cos\varphi\sin\eta_n
    -(1+\sin^2\varphi)\cos\eta_n\\
    \sin\varphi\cos\varphi\cos\eta_n
    -(1+\cos^2\varphi)\sin\eta_n\\
    -\sqrt{3}\cos(\varphi-\eta_n)
    \end{pmatrix}\, .
\end{equation}
Here \(n=0,1,2\). The omitted common displacement and arm length scale cancel
from the detector tensor.
Defining unit vectors \(\bm u\) and \(\bm v\) from spacecraft 0 toward
spacecraft 1 and 2, respectively, gives
\begin{equation}
    D_X(\varphi)
    =
    \frac{1}{2}
    \left[
    \bm u(\varphi)\otimes\bm u(\varphi)
    -
    \bm v(\varphi)\otimes\bm v(\varphi)
    \right]\, .
\end{equation}
The antenna factors are contractions of \(D_X\) with the usual plus and cross
polarization tensors. Because \(\bm u\) and \(\bm v\) are unit vectors, \(D_X\) is traceless;
their \(60^\circ\) separation gives \(D_X:D_X=3/8\) at every orbital
phase. The sky- and polarization-averaged response for either polarization
is therefore
\begin{equation}
    \frac{2}{5}D_X:D_X
    =
    {\cal R}_{\rm LW}^{X}
    =
    \frac{3}{20}\, ,
\end{equation}
where the colon denotes double contraction. The orbital average in
Eq.~\eqref{eq:response_normalized_source_power} is evaluated at
\(N_\varphi=12\) equally spaced phases
\(\varphi_m=2\pi m/N_\varphi\). This average is used only for the catalog
antenna power; no time-dependent likelihood is constructed, and Doppler
modulation is not included. Repeating the catalog power average with six
equally spaced phases gives the all-bin \(\alpha=0\) value \(I_0=1.1367\),
compared with \(1.1361\) for 12 phases; the change of \(6\times10^{-4}\)
indicates numerical stability with respect to the adopted orbital-phase
sampling.

For any source subset \(a\), the powers \(p_i\), normalized by the response, give
\begin{equation}
    S_{2,k}^{(a)}
    =
    \sum_{i\in k,a}p_i^2\, .
\end{equation}
The corresponding sums of source powers for the residual and injected
populations are \(R_k\) and \(P_{{\rm inj},k}\), respectively. For residual
sources,
\begin{equation}
    C_{2,k}^{\rm res}
    =
    \frac{S_{2,k}^{\rm res}}
         {R_k^2}\, ,
\end{equation}
for \(R_k>0\). The analysis contains 560 bins, of which 487 have nonzero
catalog residual power.

\subsection{Instrumental noise and covariance of Fourier powers}

The one-sided instrumental noise spectral density uses the optical metrology
and acceleration noise levels adopted for the LISA sensitivity
curve~\cite{Robson2019Sensitivity}, divided by the single-Michelson
long-wavelength response:
\begin{equation}
    S_{\rm inst}(f)
    =
    \frac{20}{3L^2}
    \left[
    P_{\rm OMS}(f)
    +
    \frac{4P_{\rm acc}(f)}{(2\pi f)^4}
    \right]\, ,
    \label{eq:single_x_noise_model}
\end{equation}
where \(L=2.5\times10^9\,{\rm m}\), and
\begin{align}
    P_{\rm OMS}(f)
    &=
    (1.5\times10^{-11}\,{\rm m})^2\,{\rm Hz}^{-1}
    \notag\\
    &\quad\times
    \left[
    1+\left(\frac{2.0\times10^{-3}\,{\rm Hz}}{f}\right)^4
    \right]\, ,
    \\
    P_{\rm acc}(f)
    &=
    (3.0\times10^{-15}\,{\rm m\,s^{-2}})^2\,{\rm Hz}^{-1}
    \notag\\
    &\quad\times
    \left[
    1+\left(\frac{0.4\times10^{-3}\,{\rm Hz}}{f}\right)^2
    \right]
    \notag\\
    &\quad\times
    \left[
    1+\left(\frac{f}{8.0\times10^{-3}\,{\rm Hz}}\right)^4
    \right]\, .
\end{align}
Equation~\eqref{eq:single_x_noise_model} is the long-wavelength
single-channel form; the factor \(20/3\) is
\(1/{\cal R}_{\rm LW}^{X}\). It therefore has the same equivalent incident
strain convention as \(p_i\) and the SGWB strain spectral density.

The LDC2A observation metadata give
\(T_{\rm obs}=31\,536\,000\,{\rm s}\) and a cadence of \(5\,{\rm s}\). We use
the independent positive, non-DC, non-Nyquist Fourier frequencies
\(f_j=j/T_{\rm obs}\), but not the time-series values. The half-open analysis
bins contain 315 or 316 such frequencies. Instrumental and SGWB binned powers
are the Fourier sums
\begin{equation}
    P_{{\rm inst},k}
    =
    \delta f_{\rm F}\sum_{j\in{\cal J}_k}S_{\rm inst}(f_j)\, ,
    \qquad
    T_k(\alpha)
    =
    \delta f_{\rm F}\sum_{j\in{\cal J}_k}{\cal T}_j(\alpha)\, ,
\end{equation}
where \(\delta f_{\rm F}=1/T_{\rm obs}\) and
\({\cal T}_j(\alpha)\) is given by Eq.~\eqref{eq:sgwb_mode_template}. The
binned residual model retains only the catalog power integrated over each bin
and distributes it uniformly among the \(M_k\) Fourier
frequencies, as in Eq.~\eqref{eq:residual_psd_within_bin}.

We also consider two more concentrated distributions that retain the same total
source power.
The LDC frequency \(f_i\) and derivative \(\dot f_i\) are defined at
\(t=0\)~\cite{Baghi2022LDCZonedo}. For a source at ecliptic latitude \(b_i\),
we use the frequency interval
\begin{equation}
    \begin{aligned}
    {\cal I}_i
    &={}\bigl[
    \min(f_i,f_i+\dot f_iT_{\rm obs})-\Delta f_{{\rm D},i},
    \\
    &\hspace{1.8em}
    \max(f_i,f_i+\dot f_iT_{\rm obs})+\Delta f_{{\rm D},i}
    \bigr]\, ,
    \\
    \Delta f_{{\rm D},i}
    &=
    \frac{v_{\rm orb}}{c}f_i|\cos b_i|\, ,
    \end{aligned}
\end{equation}
where \(v_{\rm orb}=29.78\,{\rm km\,s^{-1}}\). We divide \(p_i\) equally
among the Fourier frequencies in
\({\cal I}_i\); if the interval contains no Fourier frequency, it is assigned
to the closest one. As an alternative, we assign every source power
directly to its closest Fourier frequency. Both assignments
conserve each source power but do not
reproduce coherent Doppler sidebands or spectral
leakage. Among the residual sources in the analysis band, \(1.09\%\) have
intervals that cross a boundary between adjacent analysis bins. They carry
\(1.43\%\) of the residual power. Distributing each source
power within \({\cal I}_i\) moves \(0.124\%\) of the total residual power
between analysis bins and gives nonzero power to 5 of the 73 bins that are empty
when sources are assigned only by their initial intrinsic frequency.

At the fiducial parameters, the Gaussian covariance for each residual
distribution is obtained by summing the power variance at each Fourier
frequency, as in Eq.~\eqref{eq:fourier_bandpower_covariance}. The covariance
obtained for the uniform distribution within each bin is used for all
frequency-subset results. The fixed-residual comparison uses the same residual
Fourier powers in Eq.~\eqref{eq:residual_covariance_treatments}. For the uniform
distribution, the finite-source diagonal random-phase covariance uses
\(c_j=C_{2,k}^{\rm res}\) for every \(j\in{\cal J}_k\). The instrumental
spectrum is fixed in these results.
For the three-parameter Fisher comparison reported in
Appendix~\ref{app:additional_checks}, \(\eta_{\rm fid}=1\) and \(\eta\) is
marginalized without a prior while the frequency dependence of the
instrumental spectrum remains fixed. The
\(N_{\rm bin}=6,12,24,36\) cumulative subsets contain the bins with the
largest \(|\gamma_{2,\rm tot,k}|\), evaluated for
\(\Omega_{0,\rm fid}=10^{-11}\), \(\alpha=0\), and
\(\beta_{\rm fid}=1\).

\bibliography{references}

\end{document}